







%
%
%
%

\catcode `\@=11 

\def\@version{1.4}
\def\@verdate{22nd Feb 1994}

%
%
%
%


\newif\ifprod@font

\ifx\@typeface\undefined
  \def\@typeface{Comp. Modern}\prod@fontfalse
\else
  \prod@fonttrue 
\fi

\def\newfam{\alloc@8\fam\chardef\sixt@@n} 

\ifprod@font
\font\fiverm=mtr10 at 5pt
\font\fivebf=mtbx10 at 5pt
\font\fiveit=mtti10 at 5pt
\font\fivesl=mtsl10 at 5pt
\font\fivett=mttt10 at 5pt     \hyphenchar\fivett=-1
\font\fivecsc=mtcsc10 at 5pt
\font\fivesf=mtss10 at 5pt
\font\fivei=mtmi10 at 5pt      \skewchar\fivei='177
\font\fivemib=mtmib10 at 5pt   \skewchar\fivemib='177
\font\fivesy=mtsy10 at 5pt     \skewchar\fivesy='60
\font\fivesyb=mtbsy10 at 5pt   \skewchar\fivesyb='60

\font\sixrm=mtr10 at 6pt
\font\sixbf=mtbx10 at 6pt
\font\sixit=mtti10 at 6pt
\font\sixsl=mtsl10 at 6pt
\font\sixtt=mttt10 at 6pt      \hyphenchar\sixtt=-1
\font\sixcsc=mtcsc10 at 6pt
\font\sixsf=mtss10 at 6pt
\font\sixi=mtmi10 at 6pt       \skewchar\sixi='177
\font\sixmib=mtmib10 at 6pt    \skewchar\sixmib='177
\font\sixsy=mtsy10 at 6pt      \skewchar\sixsy='60
\font\sixsyb=mtbsy10 at 6pt    \skewchar\sixsyb='60

\font\sevenrm=mtr10 at 7pt
\font\sevenbf=mtbx10 at 7pt
\font\sevenit=mtti10 at 7pt
\font\sevensl=mtsl10 at 7pt
\font\seventt=mttt10 at 7pt     \hyphenchar\seventt=-1
\font\sevencsc=mtcsc10 at 7pt
\font\sevensf=mtss10 at 7pt
\font\seveni=mtmi10 at 7pt      \skewchar\seveni='177
\font\sevenmib=mtmib10 at 7pt   \skewchar\sevenmib='177
\font\sevensy=mtsy10 at 7pt     \skewchar\sevensy='60
\font\sevensyb=mtbsy10 at 7pt   \skewchar\sevensyb='60

\font\eightrm=mtr10 at 8pt
\font\eightbf=mtbx10 at 8pt
\font\eightit=mtti10 at 8pt
\font\eighti=mtmi10 at 8pt      \skewchar\eighti='177
\font\eightmib=mtmib10 at 8pt   \skewchar\eightmib='177
\font\eightsy=mtsy10 at 8pt     \skewchar\eightsy='60
\font\eightsyb=mtbsy10 at 8pt   \skewchar\eightsyb='60
\font\eightsl=mtsl10 at 8pt
\font\eighttt=mttt10 at 8pt     \hyphenchar\eighttt=-1
\font\eightcsc=mtcsc10 at 8pt
\font\eightsf=mtss10 at 8pt

\font\ninerm=mtr10 at 9pt
\font\ninebf=mtbx10 at 9pt
\font\nineit=mtti10 at 9pt
\font\ninei=mtmi10 at 9pt      \skewchar\ninei='177
\font\ninemib=mtmib10 at 9pt   \skewchar\ninemib='177
\font\ninesy=mtsy10 at 9pt     \skewchar\ninesy='60
\font\ninesyb=mtbsy10 at 9pt   \skewchar\ninesyb='60
\font\ninesl=mtsl10 at 9pt
\font\ninett=mttt10 at 9pt     \hyphenchar\ninett=-1
\font\ninecsc=mtcsc10 at 9pt
\font\ninesf=mtss10 at 9pt

\font\tenrm=mtr10
\font\tenbf=mtbx10
\font\tenit=mtti10
\font\teni=mtmi10		\skewchar\teni='177
\font\tenmib=mtmib10	\skewchar\tenmib='177
\font\tensy=mtsy10		\skewchar\tensy='60
\font\tensyb=mtbsy10	\skewchar\tensyb='60
\font\tenex=cmex10
\font\tensl=mtsl10
\font\tentt=mttt10		\hyphenchar\tentt=-1
\font\tencsc=mtcsc10
\font\tensf=mtss10

\font\elevenrm=mtr10 at 11pt
\font\elevenbf=mtbx10 at 11pt
\font\elevenit=mtti10 at 11pt
\font\eleveni=mtmi10 at 11pt      \skewchar\eleveni='177
\font\elevenmib=mtmib10 at 11pt   \skewchar\elevenmib='177
\font\elevensy=mtsy10 at 11pt     \skewchar\elevensy='60
\font\elevensyb=mtbsy10 at 11pt   \skewchar\elevensyb='60
\font\elevensl=mtsl10 at 11pt
\font\eleventt=mttt10 at 11pt     \hyphenchar\eleventt=-1
\font\elevencsc=mtcsc10 at 11pt
\font\elevensf=mtss10 at 11pt

\font\twelverm=mtr10 at 12pt
\font\twelvebf=mtbx10 at 12pt
\font\twelveit=mtti10 at 12pt
\font\twelvesl=mtsl10 at 12pt
\font\twelvett=mttt10 at 12pt     \hyphenchar\twelvett=-1
\font\twelvecsc=mtcsc10 at 12pt
\font\twelvesf=mtss10 at 12pt
\font\twelvei=mtmi10 at 12pt      \skewchar\twelvei='177
\font\twelvemib=mtmib10 at 12pt   \skewchar\twelvemib='177
\font\twelvesy=mtsy10 at 12pt     \skewchar\twelvesy='60
\font\twelvesyb=mtbsy10 at 12pt   \skewchar\twelvesyb='60

\font\fourteenrm=mtr10 at 14pt
\font\fourteenbf=mtbx10 at 14pt
\font\fourteenit=mtti10 at 14pt
\font\fourteeni=mtmi10 at 14pt      \skewchar\fourteeni='177
\font\fourteenmib=mtmib10 at 14pt   \skewchar\fourteenmib='177
\font\fourteensy=mtsy10 at 14pt     \skewchar\fourteensy='60
\font\fourteensyb=mtbsy10 at 14pt   \skewchar\fourteensyb='60
\font\fourteensl=mtsl10 at 14pt
\font\fourteentt=mttt10 at 14pt     \hyphenchar\fourteentt=-1
\font\fourteencsc=mtcsc10 at 14pt
\font\fourteensf=mtss10 at 14pt

\font\seventeenrm=mtr10 at 17pt
\font\seventeenbf=mtbx10 at 17pt
\font\seventeenit=mtti10 at 17pt
\font\seventeeni=mtmi10 at 17pt      \skewchar\seventeeni='177
\font\seventeenmib=mtmib10 at 17pt   \skewchar\seventeenmib='177
\font\seventeensy=mtsy10 at 17pt     \skewchar\seventeensy='60
\font\seventeensyb=mtbsy10 at 17pt   \skewchar\seventeensyb='60
\font\seventeensl=mtsl10 at 17pt
\font\seventeentt=mttt10 at 17pt     \hyphenchar\seventeentt=-1
\font\seventeencsc=mtcsc10 at 17pt
\font\seventeensf=mtss10 at 17pt


\newfam\xmfam
\newfam\ymfam

\font\fivexm=mtxm10 at 5pt
\font\sixxm=mtxm10 at 6pt
\font\sevenxm=mtxm10 at 7pt
\font\eightxm=mtxm10 at 8pt
\font\ninexm=mtxm10 at 9pt
\font\tenxm=mtxm10
\font\elevenxm=mtxm10 at 11pt
\font\twelvexm=mtxm10 at 12pt
\font\fourteenxm=mtxm10 at 14pt
\font\seventeenxm=mtxm10 at 17pt

\font\fiveym=mtym10 at 5pt
\font\sixym=mtym10 at 6pt
\font\sevenym=mtym10 at 7pt
\font\eightym=mtym10 at 8pt
\font\nineym=mtym10 at 9pt
\font\tenym=mtym10
\font\elevenym=mtym10 at 11pt
\font\twelveym=mtym10 at 12pt
\font\fourteenym=mtym10 at 14pt
\font\seventeenym=mtym10 at 17pt
\else
\font\fiverm=cmr5
\font\fivei=cmmi5             \skewchar\fivei='177
\font\fivemib=cmmib10 at 5pt  \skewchar\fivemib='177
\font\fivesy=cmsy5            \skewchar\fivesy='60
\font\fivesyb=cmbsy10 at 5pt  \skewchar\fivesyb='60
\font\fivebf=cmbx5

\font\sixrm=cmr6
\font\sixi=cmmi6             \skewchar\sixi='177
\font\sixmib=cmmib10 at 6pt  \skewchar\sixmib='177
\font\sixsy=cmsy6            \skewchar\sixsy='60
\font\sixsyb=cmbsy10 at 6pt  \skewchar\sixsyb='60
\font\sixbf=cmbx6

\font\sevenrm=cmr7
\font\seveni=cmmi7             \skewchar\seveni='177
\font\sevenmib=cmmib10 at 7pt  \skewchar\sevenmib='177
\font\sevensy=cmsy7            \skewchar\sevensy='60
\font\sevensyb=cmbsy10 at 7pt  \skewchar\sevensyb='60
\font\sevenbf=cmbx7

\font\eightrm=cmr8
\font\eightbf=cmbx8
\font\eightit=cmti8
\font\eighti=cmmi8			\skewchar\eighti='177
\font\eightmib=cmmib10 at 8pt	\skewchar\eightmib='177
\font\eightsy=cmsy8			\skewchar\eightsy='60
\font\eightsyb=cmbsy10 at 8pt	\skewchar\eightsyb='60
\font\eightsl=cmsl8
\font\eighttt=cmtt8			\hyphenchar\eighttt=-1
\font\eightcsc=cmcsc10 at 8pt
\font\eightsf=cmss8

\font\ninerm=cmr9
\font\ninebf=cmbx9
\font\nineit=cmti9
\font\ninei=cmmi9			\skewchar\ninei='177
\font\ninemib=cmmib10 at 9pt	\skewchar\ninemib='177
\font\ninesy=cmsy9			\skewchar\ninesy='60
\font\ninesyb=cmbsy10 at 9pt	\skewchar\ninesyb='60
\font\ninesl=cmsl9
\font\ninett=cmtt9			\hyphenchar\ninett=-1
\font\ninecsc=cmcsc10 at 9pt
\font\ninesf=cmss9

\font\tenrm=cmr10
\font\tenbf=cmbx10
\font\tenit=cmti10
\font\teni=cmmi10		\skewchar\teni='177
\font\tenmib=cmmib10	\skewchar\tenmib='177
\font\tensy=cmsy10		\skewchar\tensy='60
\font\tensyb=cmbsy10	\skewchar\tensyb='60
\font\tenex=cmex10
\font\tensl=cmsl10
\font\tentt=cmtt10		\hyphenchar\tentt=-1
\font\tencsc=cmcsc10
\font\tensf=cmss10

\font\elevenrm=cmr10 scaled \magstephalf
\font\elevenbf=cmbx10 scaled \magstephalf
\font\elevenit=cmti10 scaled \magstephalf
\font\eleveni=cmmi10 scaled \magstephalf	\skewchar\eleveni='177
\font\elevenmib=cmmib10 scaled \magstephalf	\skewchar\elevenmib='177
\font\elevensy=cmsy10 scaled \magstephalf	\skewchar\elevensy='60
\font\elevensyb=cmbsy10 scaled \magstephalf	\skewchar\elevensyb='60
\font\elevensl=cmsl10 scaled \magstephalf
\font\eleventt=cmtt10 scaled \magstephalf	\hyphenchar\eleventt=-1
\font\elevencsc=cmcsc10 scaled \magstephalf
\font\elevensf=cmss10 scaled \magstephalf

\font\twelverm=cmr10 scaled \magstep1
\font\twelvebf=cmbx10 scaled \magstep1
\font\twelvei=cmmi10 scaled \magstep1      \skewchar\twelvei='177
\font\twelvemib=cmmib10 scaled \magstep1   \skewchar\twelvemib='177
\font\twelvesy=cmsy10 scaled \magstep1     \skewchar\twelvesy='60
\font\twelvesyb=cmbsy10 scaled \magstep1   \skewchar\twelvesyb='60

\font\fourteenrm=cmr10 scaled \magstep2
\font\fourteenbf=cmbx10 scaled \magstep2
\font\fourteenit=cmti10 scaled \magstep2
\font\fourteeni=cmmi10 scaled \magstep2		\skewchar\fourteeni='177
\font\fourteenmib=cmmib10 scaled \magstep2	\skewchar\fourteenmib='177
\font\fourteensy=cmsy10 scaled \magstep2	\skewchar\fourteensy='60
\font\fourteensyb=cmbsy10 scaled \magstep2	\skewchar\fourteensyb='60
\font\fourteensl=cmsl10 scaled \magstep2
\font\fourteentt=cmtt10 scaled \magstep2	\hyphenchar\fourteentt=-1
\font\fourteencsc=cmcsc10 scaled \magstep2
\font\fourteensf=cmss10 scaled \magstep2

\font\seventeenrm=cmr10 scaled \magstep3
\font\seventeenbf=cmbx10 scaled \magstep3
\font\seventeenit=cmti10 scaled \magstep3
\font\seventeeni=cmmi10 scaled \magstep3	\skewchar\seventeeni='177
\font\seventeenmib=cmmib10 scaled \magstep3	\skewchar\seventeenmib='177
\font\seventeensy=cmsy10 scaled \magstep3	\skewchar\seventeensy='60
\font\seventeensyb=cmbsy10 scaled \magstep3	\skewchar\seventeensyb='60
\font\seventeensl=cmsl10 scaled \magstep3
\font\seventeentt=cmtt10 scaled \magstep3	\hyphenchar\seventeentt=-1
\font\seventeencsc=cmcsc10 scaled \magstep3
\font\seventeensf=cmss10 scaled \magstep3
\fi

\def\hexnumber#1{\ifcase#1 0\or1\or2\or3\or4\or5\or6\or7\or8\or9\or
  A\or B\or C\or D\or E\or F\fi}

\ifprod@font
  \edef\@xm{\hexnumber\xmfam}
  \edef\@ym{\hexnumber\ymfam}
\fi

\def\makestrut{%
  \setbox\strutbox=\hbox{%
    \vrule height.7\baselineskip depth.3\baselineskip width \z@}%
}

\def\baselinestretch{1}
\newskip\tmp@bls

\def\b@ls#1{
  \tmp@bls=#1\relax
  \baselineskip=#1\relax\makestrut
  \normalbaselineskip=\baselinestretch\tmp@bls
  \normalbaselines
}

\def\nostb@ls#1{
  \normalbaselineskip=#1\relax
  \normalbaselines
  \makestrut
}

%

\newfam\mibfam 
\newfam\sybfam 
\newfam\scfam  
\newfam\sffam  

\def\mit{\fam\@ne}

\def\cal{\fam\tw@}

\def\em{\ifdim\fontdimen1\font>\z@ \rm\else\it\fi}

\textfont3=\tenex
\scriptfont3=\tenex
\scriptscriptfont3=\tenex

\setbox0=\hbox{\tenex B} \p@renwd=\wd0 

\def\eightpoint{
  \def\rm{\fam0\eightrm}%
  \textfont0=\eightrm \scriptfont0=\sixrm \scriptscriptfont0=\fiverm%
  \textfont1=\eighti  \scriptfont1=\sixi  \scriptscriptfont1=\fivei%
  \textfont2=\eightsy \scriptfont2=\sixsy \scriptscriptfont2=\fivesy%
  \textfont\itfam=\eightit\def\it{\fam\itfam\eightit}%
  \ifprod@font
    \scriptfont\itfam=\sixit
      \scriptscriptfont\itfam=\fiveit
  \else
    \scriptfont\itfam=\eightit
      \scriptscriptfont\itfam=\eightit
  \fi
  \textfont\bffam=\eightbf%
    \scriptfont\bffam=\sixbf%
      \scriptscriptfont\bffam=\fivebf%
  \def\bf{\fam\bffam\eightbf}%
  \textfont\slfam=\eightsl\def\sl{\fam\slfam\eightsl}%
  \ifprod@font
    \scriptfont\slfam=\sixsl
      \scriptscriptfont\slfam=\fivesl
  \else
    \scriptfont\slfam=\eightsl
      \scriptscriptfont\slfam=\eightsl
  \fi
  \textfont\ttfam=\eighttt\def\tt{\fam\ttfam\eighttt}%
  \ifprod@font
    \scriptfont\ttfam=\sixtt
      \scriptscriptfont\ttfam=\fivett
  \else
    \scriptfont\ttfam=\eighttt
      \scriptscriptfont\ttfam=\eighttt
  \fi
  \textfont\scfam=\eightcsc\def\sc{\fam\scfam\eightcsc}%
  \ifprod@font
    \scriptfont\scfam=\sixcsc
      \scriptscriptfont\scfam=\fivecsc
  \else
    \scriptfont\scfam=\eightcsc
      \scriptscriptfont\scfam=\eightcsc
  \fi
  \textfont\sffam=\eightsf\def\sf{\fam\sffam\eightsf}%
  \ifprod@font
    \scriptfont\sffam=\sixsf
      \scriptscriptfont\sffam=\fivesf
  \else
    \scriptfont\sffam=\eightsf
      \scriptscriptfont\sffam=\eightsf
  \fi
  \textfont\mibfam=\eightmib
    \scriptfont\mibfam=\sixmib
      \scriptscriptfont\mibfam=\fivemib
  \textfont\sybfam=\eightsyb
    \scriptfont\sybfam=\sixsyb
      \scriptscriptfont\sybfam=\fivesyb
  \ifprod@font
    \textfont\xmfam=\eightxm
      \scriptfont\xmfam=\sixxm
        \scriptscriptfont\xmfam=\fivexm
    \textfont\ymfam=\eightym
      \scriptfont\ymfam=\sixym
        \scriptscriptfont\ymfam=\fiveym
  \fi
  \def\oldstyle{\fam\@ne\eighti}%
  \def\boldstyle{\fam\mibfam\eightmib}%
  \b@ls{10pt}\rm%
}

\def\ninepoint{
  \def\rm{\fam0\ninerm}%
  \textfont0=\ninerm \scriptfont0=\sixrm \scriptscriptfont0=\fiverm%
  \textfont1=\ninei  \scriptfont1=\sixi  \scriptscriptfont1=\fivei%
  \textfont2=\ninesy \scriptfont2=\sixsy \scriptscriptfont2=\fivesy%
  \textfont\itfam=\nineit\def\it{\fam\itfam\nineit}%
  \ifprod@font
    \scriptfont\itfam=\sixit
      \scriptscriptfont\itfam=\fiveit
  \else
    \scriptfont\itfam=\nineit
      \scriptscriptfont\itfam=\nineit
  \fi
  \textfont\bffam=\ninebf%
    \scriptfont\bffam=\sixbf%
      \scriptscriptfont\bffam=\fivebf%
  \def\bf{\fam\bffam\ninebf}%
  \textfont\slfam=\ninesl\def\sl{\fam\slfam\ninesl}%
  \ifprod@font
    \scriptfont\slfam=\sixsl
      \scriptscriptfont\slfam=\fivesl
  \else
    \scriptfont\slfam=\ninesl
      \scriptscriptfont\slfam=\ninesl
  \fi
  \textfont\ttfam=\ninett\def\tt{\fam\ttfam\ninett}%
  \ifprod@font
    \scriptfont\ttfam=\sixtt
      \scriptscriptfont\ttfam=\fivett
  \else
    \scriptfont\ttfam=\ninett
      \scriptscriptfont\ttfam=\ninett
  \fi
  \textfont\scfam=\ninecsc\def\sc{\fam\scfam\ninecsc}%
  \ifprod@font
    \scriptfont\scfam=\sixcsc
      \scriptscriptfont\scfam=\fivecsc
  \else
    \scriptfont\scfam=\ninecsc
      \scriptscriptfont\scfam=\ninecsc
  \fi
  \textfont\sffam=\ninesf\def\sf{\fam\sffam\ninesf}%
  \ifprod@font
    \scriptfont\sffam=\sixsf
      \scriptscriptfont\sffam=\fivesf
  \else
    \scriptfont\sffam=\ninesf
      \scriptscriptfont\sffam=\ninesf
  \fi
  \textfont\mibfam=\ninemib
    \scriptfont\mibfam=\sixmib
      \scriptscriptfont\mibfam=\fivemib
  \textfont\sybfam=\ninesyb
    \scriptfont\sybfam=\sixsyb
      \scriptscriptfont\sybfam=\fivesyb
  \ifprod@font
    \textfont\xmfam=\ninexm
      \scriptfont\xmfam=\sixxm
        \scriptscriptfont\xmfam=\fivexm
    \textfont\ymfam=\nineym
      \scriptfont\ymfam=\sixym
        \scriptscriptfont\ymfam=\fiveym
  \fi
  \def\oldstyle{\fam\@ne\ninei}%
  \def\boldstyle{\fam\mibfam\ninemib}%
  \b@ls{\TextLeading plus \Feathering}\rm%
}

\def\tenpoint{
  \def\rm{\fam0\tenrm}%
  \textfont0=\tenrm \scriptfont0=\sevenrm \scriptscriptfont0=\fiverm%
  \textfont1=\teni  \scriptfont1=\seveni  \scriptscriptfont1=\fivei%
  \textfont2=\tensy \scriptfont2=\sevensy \scriptscriptfont2=\fivesy%
  \textfont\itfam=\tenit\def\it{\fam\itfam\tenit}%
  \ifprod@font
    \scriptfont\itfam=\sevenit
      \scriptscriptfont\itfam=\fiveit
  \else
    \scriptfont\itfam=\tenit
      \scriptscriptfont\itfam=\tenit
  \fi
  \textfont\bffam=\tenbf%
    \scriptfont\bffam=\sevenbf%
      \scriptscriptfont\bffam=\fivebf%
  \def\bf{\fam\bffam\tenbf}%
  \textfont\slfam=\tensl\def\sl{\fam\slfam\tensl}%
  \ifprod@font
    \scriptfont\slfam=\sevensl
      \scriptscriptfont\slfam=\fivesl
  \else
    \scriptfont\slfam=\tensl
      \scriptscriptfont\slfam=\tensl
  \fi
  \textfont\ttfam=\tentt\def\tt{\fam\ttfam\tentt}%
  \ifprod@font
    \scriptfont\ttfam=\seventt
      \scriptscriptfont\ttfam=\fivett
  \else
    \scriptfont\ttfam=\tentt
      \scriptscriptfont\ttfam=\tentt
  \fi
  \textfont\scfam=\tencsc\def\sc{\fam\scfam\tencsc}%
  \ifprod@font
    \scriptfont\scfam=\sevencsc
      \scriptscriptfont\scfam=\fivecsc
  \else
    \scriptfont\scfam=\tencsc
      \scriptscriptfont\scfam=\tencsc
  \fi
  \textfont\sffam=\tensf\def\sf{\fam\sffam\tensf}%
  \ifprod@font
    \scriptfont\sffam=\sevensf
      \scriptscriptfont\sffam=\fivesf
  \else
    \scriptfont\sffam=\tensf
      \scriptscriptfont\sffam=\tensf
  \fi
  \textfont\mibfam=\tenmib
    \scriptfont\mibfam=\sevenmib
      \scriptscriptfont\mibfam=\fivemib
  \textfont\sybfam=\tensyb
    \scriptfont\sybfam=\sevensyb
      \scriptscriptfont\sybfam=\fivesyb
  \ifprod@font
    \textfont\xmfam=\tenxm
      \scriptfont\xmfam=\sevenxm
        \scriptscriptfont\xmfam=\fivexm
    \textfont\ymfam=\tenym
      \scriptfont\ymfam=\sevenym
        \scriptscriptfont\ymfam=\fiveym
  \fi
  \def\oldstyle{\fam\@ne\teni}%
  \def\boldstyle{\fam\mibfam\tenmib}%
  \b@ls{11pt}\rm%
}

\def\elevenpoint{
  \def\rm{\fam0\elevenrm}%
  \textfont0=\elevenrm \scriptfont0=\eightrm \scriptscriptfont0=\sixrm%
  \textfont1=\eleveni  \scriptfont1=\eighti  \scriptscriptfont1=\sixi%
  \textfont2=\elevensy \scriptfont2=\eightsy \scriptscriptfont2=\sixsy%
  \textfont\itfam=\elevenit\def\it{\fam\itfam\elevenit}%
  \ifprod@font
    \scriptfont\itfam=\eightit
      \scriptscriptfont\itfam=\sixit
  \else
    \scriptfont\itfam=\elevenit
      \scriptscriptfont\itfam=\elevenit
  \fi
  \textfont\bffam=\elevenbf%
    \scriptfont\bffam=\eightbf%
      \scriptscriptfont\bffam=\sixbf%
  \def\bf{\fam\bffam\elevenbf}%
  \textfont\slfam=\elevensl\def\sl{\fam\slfam\elevensl}%
  \ifprod@font
    \scriptfont\slfam=\eightsl
      \scriptscriptfont\slfam=\sixsl
  \else
    \scriptfont\slfam=\elevensl
      \scriptscriptfont\slfam=\elevensl
  \fi
  \textfont\ttfam=\eleventt\def\tt{\fam\ttfam\eleventt}%
  \ifprod@font
    \scriptfont\ttfam=\eighttt
      \scriptscriptfont\ttfam=\sixtt
  \else
    \scriptfont\ttfam=\eleventt
      \scriptscriptfont\ttfam=\eleventt
  \fi
  \textfont\scfam=\elevencsc\def\sc{\fam\scfam\elevencsc}%
  \ifprod@font
    \scriptfont\scfam=\eightcsc
      \scriptscriptfont\scfam=\sixcsc
  \else
    \scriptfont\scfam=\elevencsc
      \scriptscriptfont\scfam=\elevencsc
  \fi
  \textfont\sffam=\elevensf\def\sf{\fam\sffam\elevensf}%
  \ifprod@font
    \scriptfont\sffam=\eightsf
      \scriptscriptfont\sffam=\sixsf
  \else
    \scriptfont\sffam=\elevensf
      \scriptscriptfont\sffam=\elevensf
  \fi
  \textfont\mibfam=\elevenmib
    \scriptfont\mibfam=\eightmib
      \scriptscriptfont\mibfam=\sixmib
  \textfont\sybfam=\elevensyb
    \scriptfont\sybfam=\eightsyb
      \scriptscriptfont\sybfam=\sixsyb
  \ifprod@font
    \textfont\xmfam=\elevenxm
      \scriptfont\xmfam=\eightxm
       \scriptscriptfont\xmfam=\sixxm
    \textfont\ymfam=\elevenym
      \scriptfont\ymfam=\eightym
        \scriptscriptfont\ymfam=\sixym
   \fi
  \def\oldstyle{\fam\@ne\eleveni}%
  \def\boldstyle{\fam\mibfam\elevenmib}%
  \b@ls{13pt}\rm%
}

\def\fourteenpoint{
  \def\rm{\fam0\fourteenrm}%
  \textfont0\fourteenrm  \scriptfont0\tenrm  \scriptscriptfont0\sevenrm%
  \textfont1\fourteeni   \scriptfont1\teni   \scriptscriptfont1\seveni%
  \textfont2\fourteensy  \scriptfont2\tensy  \scriptscriptfont2\sevensy%
  \textfont\itfam=\fourteenit\def\it{\fam\itfam\fourteenit}%
  \ifprod@font
    \scriptfont\itfam=\tenit
      \scriptscriptfont\itfam=\sevenit
  \else
    \scriptfont\itfam=\fourteenit
      \scriptscriptfont\itfam=\fourteenit
  \fi
  \textfont\bffam=\fourteenbf%
    \scriptfont\bffam=\tenbf%
      \scriptscriptfont\bffam=\sevenbf%
  \def\bf{\fam\bffam\fourteenbf}%
  \textfont\slfam=\fourteensl\def\sl{\fam\slfam\fourteensl}%
  \ifprod@font
    \scriptfont\slfam=\tensl
      \scriptscriptfont\slfam=\sevensl
  \else
    \scriptfont\slfam=\fourteensl
      \scriptscriptfont\slfam=\fourteensl
  \fi
  \textfont\ttfam=\fourteentt\def\tt{\fam\ttfam\fourteentt}%
  \ifprod@font
    \scriptfont\ttfam=\tentt
      \scriptscriptfont\ttfam=\seventt
  \else
    \scriptfont\ttfam=\fourteentt
      \scriptscriptfont\ttfam=\fourteentt
  \fi
  \textfont\scfam=\fourteencsc\def\sc{\fam\scfam\fourteencsc}%
  \ifprod@font
    \scriptfont\scfam=\tencsc
      \scriptscriptfont\scfam=\sevencsc
  \else
    \scriptfont\scfam=\fourteencsc
      \scriptscriptfont\scfam=\fourteencsc
  \fi
  \textfont\sffam=\fourteensf\def\sf{\fam\sffam\fourteensf}%
  \ifprod@font
    \scriptfont\sffam=\tensf
      \scriptscriptfont\sffam=\sevensf
  \else
    \scriptfont\sffam=\fourteensf
      \scriptscriptfont\sffam=\fourteensf
  \fi
  \textfont\mibfam=\fourteenmib
    \scriptfont\mibfam=\tenmib
      \scriptscriptfont\mibfam=\sevenmib
  \textfont\sybfam=\fourteensyb
    \scriptfont\sybfam=\tensyb
      \scriptscriptfont\sybfam=\sevensyb
  \ifprod@font
    \textfont\xmfam=\fourteenxm
      \scriptfont\xmfam=\tenxm
        \scriptscriptfont\xmfam=\sevenxm
   \textfont\ymfam=\fourteenym
      \scriptfont\ymfam=\tenym
        \scriptscriptfont\ymfam=\sevenym
  \fi
  \def\oldstyle{\fam\@ne\fourteeni}%
  \def\boldstyle{\fam\mibfam\fourteenmib}%
  \b@ls{17pt}\rm%
}

\def\seventeenpoint{
  \def\rm{\fam0\seventeenrm}%
  \textfont0\seventeenrm  \scriptfont0\twelverm  \scriptscriptfont0\tenrm%
  \textfont1\seventeeni   \scriptfont1\twelvei   \scriptscriptfont1\teni%
  \textfont2\seventeensy  \scriptfont2\twelvesy  \scriptscriptfont2\tensy%
  \textfont\itfam=\seventeenit\def\it{\fam\itfam\seventeenit}%
  \ifprod@font
    \scriptfont\itfam=\twelveit
      \scriptscriptfont\itfam=\tenit
  \else
    \scriptfont\itfam=\seventeenit
      \scriptscriptfont\itfam=\seventeenit
  \fi
  \textfont\bffam=\seventeenbf%
    \scriptfont\bffam=\twelvebf%
      \scriptscriptfont\bffam=\tenbf%
  \def\bf{\fam\bffam\seventeenbf}%
  \textfont\slfam=\seventeensl\def\sl{\fam\slfam\seventeensl}%
  \ifprod@font
    \scriptfont\slfam=\twelvesl
      \scriptscriptfont\slfam=\tensl
  \else
    \scriptfont\slfam=\seventeensl
      \scriptscriptfont\slfam=\seventeensl
  \fi
  \textfont\ttfam=\seventeentt\def\tt{\fam\ttfam\seventeentt}%
  \ifprod@font
    \scriptfont\ttfam=\twelvett
      \scriptscriptfont\ttfam=\tentt
  \else
    \scriptfont\ttfam=\seventeentt
      \scriptscriptfont\ttfam=\seventeentt
  \fi
  \textfont\scfam=\seventeencsc\def\sc{\fam\scfam\seventeencsc}%
  \ifprod@font
    \scriptfont\scfam=\twelvecsc
      \scriptscriptfont\scfam=\tencsc
  \else
    \scriptfont\scfam=\seventeencsc
      \scriptscriptfont\scfam=\seventeencsc
  \fi
  \textfont\sffam=\seventeensf\def\sf{\fam\sffam\seventeensf}%
  \ifprod@font
    \scriptfont\sffam=\twelvesf
      \scriptscriptfont\sffam=\tensf
  \else
    \scriptfont\sffam=\seventeensf
      \scriptscriptfont\sffam=\seventeensf
  \fi
  \textfont\mibfam=\seventeenmib
    \scriptfont\mibfam=\twelvemib
      \scriptscriptfont\mibfam=\tenmib
  \textfont\sybfam=\seventeensyb
    \scriptfont\sybfam=\twelvesyb
      \scriptscriptfont\sybfam=\tensyb
  \ifprod@font
    \textfont\xmfam=\seventeenxm
      \scriptfont\xmfam=\twelvexm
        \scriptscriptfont\xmfam=\tenxm
    \textfont\ymfam=\seventeenym
      \scriptfont\ymfam=\twelveym
        \scriptscriptfont\ymfam=\tenym
  \fi
  \def\oldstyle{\fam\@ne\seventeeni}%
  \def\boldstyle{\fam\mibfam\seventeenmib}%
  \b@ls{20pt}\rm%
}

\lineskip=1pt      \normallineskip=\lineskip
\lineskiplimit=\z@ \normallineskiplimit=\lineskiplimit



\def\la{\mathrel{\mathchoice {\vcenter{\offinterlineskip\halign{\hfil
$\displaystyle##$\hfil\cr<\cr\sim\cr}}}
{\vcenter{\offinterlineskip\halign{\hfil$\textstyle##$\hfil\cr
<\cr\sim\cr}}}
{\vcenter{\offinterlineskip\halign{\hfil$\scriptstyle##$\hfil\cr
<\cr\sim\cr}}}
{\vcenter{\offinterlineskip\halign{\hfil$\scriptscriptstyle##$\hfil\cr
<\cr\sim\cr}}}}}

\def\ga{\mathrel{\mathchoice {\vcenter{\offinterlineskip\halign{\hfil
$\displaystyle##$\hfil\cr>\cr\sim\cr}}}
{\vcenter{\offinterlineskip\halign{\hfil$\textstyle##$\hfil\cr
>\cr\sim\cr}}}
{\vcenter{\offinterlineskip\halign{\hfil$\scriptstyle##$\hfil\cr
>\cr\sim\cr}}}
{\vcenter{\offinterlineskip\halign{\hfil$\scriptscriptstyle##$\hfil\cr
>\cr\sim\cr}}}}}

\def\getsto{\mathrel{\mathchoice {\vcenter{\offinterlineskip
\halign{\hfil
$\displaystyle##$\hfil\cr\gets\cr\to\cr}}}
{\vcenter{\offinterlineskip\halign{\hfil$\textstyle##$\hfil\cr\gets
\cr\to\cr}}}
{\vcenter{\offinterlineskip\halign{\hfil$\scriptstyle##$\hfil\cr\gets
\cr\to\cr}}}
{\vcenter{\offinterlineskip\halign{\hfil$\scriptscriptstyle##$\hfil\cr
\gets\cr\to\cr}}}}}

\def\lid{\mathrel{\mathchoice {\vcenter{\offinterlineskip\halign{\hfil
$\displaystyle##$\hfil\cr<\cr\noalign{\vskip1.2pt}=\cr}}}
{\vcenter{\offinterlineskip\halign{\hfil$\textstyle##$\hfil\cr<\cr
\noalign{\vskip1.2pt}=\cr}}}
{\vcenter{\offinterlineskip\halign{\hfil$\scriptstyle##$\hfil\cr<\cr
\noalign{\vskip1pt}=\cr}}}
{\vcenter{\offinterlineskip\halign{\hfil$\scriptscriptstyle##$\hfil\cr
<\cr
\noalign{\vskip0.9pt}=\cr}}}}}

\def\gid{\mathrel{\mathchoice {\vcenter{\offinterlineskip\halign{\hfil
$\displaystyle##$\hfil\cr>\cr\noalign{\vskip1.2pt}=\cr}}}
{\vcenter{\offinterlineskip\halign{\hfil$\textstyle##$\hfil\cr>\cr
\noalign{\vskip1.2pt}=\cr}}}
{\vcenter{\offinterlineskip\halign{\hfil$\scriptstyle##$\hfil\cr>\cr
\noalign{\vskip1pt}=\cr}}}
{\vcenter{\offinterlineskip\halign{\hfil$\scriptscriptstyle##$\hfil\cr
>\cr
\noalign{\vskip0.9pt}=\cr}}}}}

\def\grole{\mathrel{\mathchoice {\vcenter{\offinterlineskip\halign{\hfil
$\displaystyle##$\hfil\cr>\cr\noalign{\vskip-1.5pt}<\cr}}}
{\vcenter{\offinterlineskip\halign{\hfil$\textstyle##$\hfil\cr
>\cr\noalign{\vskip-1.5pt}<\cr}}}
{\vcenter{\offinterlineskip\halign{\hfil$\scriptstyle##$\hfil\cr
>\cr\noalign{\vskip-1pt}<\cr}}}
{\vcenter{\offinterlineskip\halign{\hfil$\scriptscriptstyle##$\hfil\cr
>\cr\noalign{\vskip-0.5pt}<\cr}}}}}

\def\leogr{\mathrel{\mathchoice {\vcenter{\offinterlineskip\halign{\hfil
$\displaystyle##$\hfil\cr<\cr\noalign{\vskip-1.5pt}>\cr}}}
{\vcenter{\offinterlineskip\halign{\hfil$\textstyle##$\hfil\cr
<\cr\noalign{\vskip-1.5pt}>\cr}}}
{\vcenter{\offinterlineskip\halign{\hfil$\scriptstyle##$\hfil\cr
<\cr\noalign{\vskip-1pt}>\cr}}}
{\vcenter{\offinterlineskip\halign{\hfil$\scriptscriptstyle##$\hfil\cr
<\cr\noalign{\vskip-0.5pt}>\cr}}}}}

\def\loa{\mathrel{\mathchoice {\vcenter{\offinterlineskip\halign{\hfil
$\displaystyle##$\hfil\cr<\cr\approx\cr}}}
{\vcenter{\offinterlineskip\halign{\hfil$\textstyle##$\hfil\cr
<\cr\approx\cr}}}
{\vcenter{\offinterlineskip\halign{\hfil$\scriptstyle##$\hfil\cr
<\cr\approx\cr}}}
{\vcenter{\offinterlineskip\halign{\hfil$\scriptscriptstyle##$\hfil\cr
<\cr\approx\cr}}}}}

\def\goa{\mathrel{\mathchoice {\vcenter{\offinterlineskip\halign{\hfil
$\displaystyle##$\hfil\cr>\cr\approx\cr}}}
{\vcenter{\offinterlineskip\halign{\hfil$\textstyle##$\hfil\cr
>\cr\approx\cr}}}
{\vcenter{\offinterlineskip\halign{\hfil$\scriptstyle##$\hfil\cr
>\cr\approx\cr}}}
{\vcenter{\offinterlineskip\halign{\hfil$\scriptscriptstyle##$\hfil\cr
>\cr\approx\cr}}}}}

\def\diameter{{\ifmmode\mathchoice
{\ooalign{\hfil\hbox{$\displaystyle/$}\hfil\crcr
{\hbox{$\displaystyle\mathchar"20D$}}}}
{\ooalign{\hfil\hbox{$\textstyle/$}\hfil\crcr
{\hbox{$\textstyle\mathchar"20D$}}}}
{\ooalign{\hfil\hbox{$\scriptstyle/$}\hfil\crcr
{\hbox{$\scriptstyle\mathchar"20D$}}}}
{\ooalign{\hfil\hbox{$\scriptscriptstyle/$}\hfil\crcr
{\hbox{$\scriptscriptstyle\mathchar"20D$}}}}
\else{\ooalign{\hfil/\hfil\crcr\mathhexbox20D}}%
\fi}}

\def\sq{\ifmmode\squareforqed\else{\unskip\nobreak\hfil
\penalty50\hskip1em\null\nobreak\hfil\squareforqed
\parfillskip=0pt\finalhyphendemerits=0\endgraf}\fi}
\def\squareforqed{\hbox{\rlap{$\sqcap$}$\sqcup$}}


\def\bbbc{{\mathchoice {\setbox0=\hbox{$\displaystyle\rm C$}\hbox{\hbox
to0pt{\kern0.4\wd0\vrule height0.9\ht0\hss}\box0}}
{\setbox0=\hbox{$\textstyle\rm C$}\hbox{\hbox
to0pt{\kern0.4\wd0\vrule height0.9\ht0\hss}\box0}}
{\setbox0=\hbox{$\scriptstyle\rm C$}\hbox{\hbox
to0pt{\kern0.4\wd0\vrule height0.9\ht0\hss}\box0}}
{\setbox0=\hbox{$\scriptscriptstyle\rm C$}\hbox{\hbox
to0pt{\kern0.4\wd0\vrule height0.9\ht0\hss}\box0}}}}
\def\bbbq{{\mathchoice {\setbox0=\hbox{$\displaystyle\rm
Q$}\hbox{\raise
0.15\ht0\hbox to0pt{\kern0.4\wd0\vrule height0.8\ht0\hss}\box0}}
{\setbox0=\hbox{$\textstyle\rm Q$}\hbox{\raise
0.15\ht0\hbox to0pt{\kern0.4\wd0\vrule height0.8\ht0\hss}\box0}}
{\setbox0=\hbox{$\scriptstyle\rm Q$}\hbox{\raise
0.15\ht0\hbox to0pt{\kern0.4\wd0\vrule height0.7\ht0\hss}\box0}}
{\setbox0=\hbox{$\scriptscriptstyle\rm Q$}\hbox{\raise
0.15\ht0\hbox to0pt{\kern0.4\wd0\vrule height0.7\ht0\hss}\box0}}}}
\def\bbbt{{\mathchoice {\setbox0=\hbox{$\displaystyle\rm
T$}\hbox{\hbox to0pt{\kern0.3\wd0\vrule height0.9\ht0\hss}\box0}}
{\setbox0=\hbox{$\textstyle\rm T$}\hbox{\hbox
to0pt{\kern0.3\wd0\vrule height0.9\ht0\hss}\box0}}
{\setbox0=\hbox{$\scriptstyle\rm T$}\hbox{\hbox
to0pt{\kern0.3\wd0\vrule height0.9\ht0\hss}\box0}}
{\setbox0=\hbox{$\scriptscriptstyle\rm T$}\hbox{\hbox
to0pt{\kern0.3\wd0\vrule height0.9\ht0\hss}\box0}}}}
\def\bbbs{{\mathchoice
{\setbox0=\hbox{$\displaystyle     \rm S$}\hbox{\raise0.5\ht0\hbox
to0pt{\kern0.35\wd0\vrule height0.45\ht0\hss}\hbox
to0pt{\kern0.55\wd0\vrule height0.5\ht0\hss}\box0}}
{\setbox0=\hbox{$\textstyle        \rm S$}\hbox{\raise0.5\ht0\hbox
to0pt{\kern0.35\wd0\vrule height0.45\ht0\hss}\hbox
to0pt{\kern0.55\wd0\vrule height0.5\ht0\hss}\box0}}
{\setbox0=\hbox{$\scriptstyle      \rm S$}\hbox{\raise0.5\ht0\hbox
to0pt{\kern0.35\wd0\vrule height0.45\ht0\hss}\raise0.05\ht0\hbox
to0pt{\kern0.5\wd0\vrule height0.45\ht0\hss}\box0}}
{\setbox0=\hbox{$\scriptscriptstyle\rm S$}\hbox{\raise0.5\ht0\hbox
to0pt{\kern0.4\wd0\vrule height0.45\ht0\hss}\raise0.05\ht0\hbox
to0pt{\kern0.55\wd0\vrule height0.45\ht0\hss}\box0}}}}
\def\bbbz{{\mathchoice {\hbox{$\sf\textstyle Z\kern-0.4em Z$}}
{\hbox{$\sf\textstyle Z\kern-0.4em Z$}}
{\hbox{$\sf\scriptstyle Z\kern-0.3em Z$}}
{\hbox{$\sf\scriptscriptstyle Z\kern-0.2em Z$}}}}


\ifprod@font
  \mathchardef\la="3\@xm2E
  \mathchardef\getsto="3\@xm1C
  \mathchardef\lid="3\@xm35
  \mathchardef\grole="3\@xm3F
  \mathchardef\loa="3\@xm2F
  \mathchardef\ga="3\@xm26
  \mathchardef\gid="3\@xm3D
  \mathchardef\leogr="3\@xm37
  \mathchardef\goa="3\@xm27
  \mathchardef\sq="0\@xm03
%
%
\def\diameter{{%
  \ifmmode
    \mathchoice
    {\ooalign{\hfil\hbox{$\displaystyle/$}\hfil\crcr
    {\lower.2ex\hbox{$\displaystyle\mathchar"20D$}}}}%
    {\ooalign{\hfil\hbox{$\textstyle/$}\hfil\crcr
    {\lower.2ex\hbox{$\textstyle\mathchar"20D$}}}}%
    {\ooalign{\hfil\hbox{$\scriptstyle/$}\hfil\crcr
    {\lower.1ex\hbox{$\scriptstyle\mathchar"20D$}}}}%
    {\ooalign{\hfil\hbox{$\scriptscriptstyle/$}\hfil\crcr
    {\lower.1ex\hbox{$\scriptscriptstyle\mathchar"20D$}}}}%
  \else
    {\ooalign{\hfil/\hfil\crcr\lower.2ex\hbox{\mathhexbox20D}}}%
  \fi
}}
%
%

\def\bbbc{{\Bbb{C}}}
\def\bbbq{{\Bbb{Q}}}
\def\bbbt{{\Bbb{T}}}
\def\bbbs{{\Bbb{S}}}
\def\bbbz{{\Bbb{Z}}}
\fi


\ifprod@font
\mathchardef\boxdot="2\@xm00
\mathchardef\boxplus="2\@xm01
\mathchardef\boxtimes="2\@xm02
\mathchardef\square="0\@xm03
\mathchardef\blacksquare="0\@xm04
\mathchardef\centerdot="2\@xm05
\mathchardef\lozenge="0\@xm06
\mathchardef\blacklozenge="0\@xm07
\mathchardef\circlearrowright="3\@xm08
\mathchardef\circlearrowleft="3\@xm09
\mathchardef\rightleftharpoons="3\@xm0A
\mathchardef\leftrightharpoons="3\@xm0B
\mathchardef\boxminus="2\@xm0C
\mathchardef\Vdash="3\@xm0D
\mathchardef\Vvdash="3\@xm0E
\mathchardef\vDash="3\@xm0F
\mathchardef\twoheadrightarrow="3\@xm10
\mathchardef\twoheadleftarrow="3\@xm11
\mathchardef\leftleftarrows="3\@xm12
\mathchardef\rightrightarrows="3\@xm13
\mathchardef\upuparrows="3\@xm14
\mathchardef\downdownarrows="3\@xm15
\mathchardef\upharpoonright="3\@xm16

\mathchardef\downharpoonright="3\@xm17
\mathchardef\upharpoonleft="3\@xm18
\mathchardef\downharpoonleft="3\@xm19
\mathchardef\rightarrowtail="3\@xm1A
\mathchardef\leftarrowtail="3\@xm1B
\mathchardef\leftrightarrows="3\@xm1C
\mathchardef\rightleftarrows="3\@xm1D
\mathchardef\Lsh="3\@xm1E
\mathchardef\Rsh="3\@xm1F
\mathchardef\rightsquigarrow="3\@xm20
\mathchardef\leftrightsquigarrow="3\@xm21
\mathchardef\looparrowleft="3\@xm22
\mathchardef\looparrowright="3\@xm23
\mathchardef\circeq="3\@xm24
\mathchardef\succsim="3\@xm25
\mathchardef\gtrsim="3\@xm26
\mathchardef\gtrapprox="3\@xm27
\mathchardef\multimap="3\@xm28
\mathchardef\therefore="3\@xm29
\mathchardef\because="3\@xm2A
\mathchardef\doteqdot="3\@xm2B

\mathchardef\triangleq="3\@xm2C
\mathchardef\precsim="3\@xm2D
\mathchardef\lesssim="3\@xm2E
\mathchardef\lessapprox="3\@xm2F
\mathchardef\eqslantless="3\@xm30
\mathchardef\eqslantgtr="3\@xm31
\mathchardef\curlyeqprec="3\@xm32
\mathchardef\curlyeqsucc="3\@xm33
\mathchardef\preccurlyeq="3\@xm34
\mathchardef\leqq="3\@xm35
\mathchardef\leqslant="3\@xm36
\mathchardef\lessgtr="3\@xm37
\mathchardef\backprime="0\@xm38
\mathchardef\risingdotseq="3\@xm3A
\mathchardef\fallingdotseq="3\@xm3B
\mathchardef\succcurlyeq="3\@xm3C
\mathchardef\geqq="3\@xm3D
\mathchardef\geqslant="3\@xm3E
\mathchardef\gtrless="3\@xm3F
\mathchardef\sqsubset="3\@xm40
\mathchardef\sqsupset="3\@xm41
\mathchardef\vartriangleright="3\@xm42
\mathchardef\vartriangleleft="3\@xm43
\mathchardef\trianglerighteq="3\@xm44
\mathchardef\trianglelefteq="3\@xm45
\mathchardef\bigstar="0\@xm46
\mathchardef\between="3\@xm47
\mathchardef\blacktriangledown="0\@xm48
\mathchardef\blacktriangleright="3\@xm49
\mathchardef\blacktriangleleft="3\@xm4A
\mathchardef\vartriangle="0\@xm4D
\mathchardef\blacktriangle="0\@xm4E
\mathchardef\triangledown="0\@xm4F
\mathchardef\eqcirc="3\@xm50
\mathchardef\lesseqgtr="3\@xm51
\mathchardef\gtreqless="3\@xm52
\mathchardef\lesseqqgtr="3\@xm53
\mathchardef\gtreqqless="3\@xm54
\mathchardef\Rrightarrow="3\@xm56
\mathchardef\Lleftarrow="3\@xm57
\mathchardef\veebar="2\@xm59
\mathchardef\barwedge="2\@xm5A
\mathchardef\doublebarwedge="2\@xm5B
\mathchardef\angle="0\@xm5C
\mathchardef\measuredangle="0\@xm5D
\mathchardef\sphericalangle="0\@xm5E
\mathchardef\varpropto="3\@xm5F
\mathchardef\smallsmile="3\@xm60
\mathchardef\smallfrown="3\@xm61
\mathchardef\Subset="3\@xm62
\mathchardef\Supset="3\@xm63
\mathchardef\Cup="2\@xm64

\mathchardef\Cap="2\@xm65

\mathchardef\curlywedge="2\@xm66
\mathchardef\curlyvee="2\@xm67
\mathchardef\leftthreetimes="2\@xm68
\mathchardef\rightthreetimes="2\@xm69
\mathchardef\subseteqq="3\@xm6A
\mathchardef\supseteqq="3\@xm6B
\mathchardef\bumpeq="3\@xm6C
\mathchardef\Bumpeq="3\@xm6D
\mathchardef\lll="3\@xm6E

\mathchardef\ggg="3\@xm6F

\mathchardef\circledS="0\@xm73
\mathchardef\pitchfork="3\@xm74
\mathchardef\dotplus="2\@xm75
\mathchardef\backsim="3\@xm76
\mathchardef\backsimeq="3\@xm77
\mathchardef\complement="0\@xm7B
\mathchardef\intercal="2\@xm7C
\mathchardef\circledcirc="2\@xm7D
\mathchardef\circledast="2\@xm7E
\mathchardef\circleddash="2\@xm7F
\def\ulcorner{\delimiter"4\@xm70\@xm70 }
\def\urcorner{\delimiter"5\@xm71\@xm71 }
\def\llcorner{\delimiter"4\@xm78\@xm78 }
\def\lrcorner{\delimiter"5\@xm79\@xm79 }
\def\yen{\mathhexbox\@xm55 }
\def\checkmark{\mathhexbox\@xm58 }
\def\circledR{\mathhexbox\@xm72 }
\def\maltese{\mathhexbox\@xm7A }
\mathchardef\lvertneqq="3\@ym00
\mathchardef\gvertneqq="3\@ym01
\mathchardef\nleq="3\@ym02
\mathchardef\ngeq="3\@ym03
\mathchardef\nless="3\@ym04
\mathchardef\ngtr="3\@ym05
\mathchardef\nprec="3\@ym06
\mathchardef\nsucc="3\@ym07
\mathchardef\lneqq="3\@ym08
\mathchardef\gneqq="3\@ym09
\mathchardef\nleqslant="3\@ym0A
\mathchardef\ngeqslant="3\@ym0B
\mathchardef\lneq="3\@ym0C
\mathchardef\gneq="3\@ym0D
\mathchardef\npreceq="3\@ym0E
\mathchardef\nsucceq="3\@ym0F
\mathchardef\precnsim="3\@ym10
\mathchardef\succnsim="3\@ym11
\mathchardef\lnsim="3\@ym12
\mathchardef\gnsim="3\@ym13
\mathchardef\nleqq="3\@ym14
\mathchardef\ngeqq="3\@ym15
\mathchardef\precneqq="3\@ym16
\mathchardef\succneqq="3\@ym17
\mathchardef\precnapprox="3\@ym18
\mathchardef\succnapprox="3\@ym19
\mathchardef\lnapprox="3\@ym1A
\mathchardef\gnapprox="3\@ym1B
\mathchardef\nsim="3\@ym1C
\mathchardef\ncong="3\@ym1D

\mathchardef\varsubsetneq="3\@ym20
\mathchardef\varsupsetneq="3\@ym21
\mathchardef\nsubseteqq="3\@ym22
\mathchardef\nsupseteqq="3\@ym23
\mathchardef\subsetneqq="3\@ym24
\mathchardef\supsetneqq="3\@ym25
\mathchardef\varsubsetneqq="3\@ym26
\mathchardef\varsupsetneqq="3\@ym27
\mathchardef\subsetneq="3\@ym28
\mathchardef\supsetneq="3\@ym29
\mathchardef\nsubseteq="3\@ym2A
\mathchardef\nsupseteq="3\@ym2B
\mathchardef\nparallel="3\@ym2C
\mathchardef\nmid="3\@ym2D
\mathchardef\nshortmid="3\@ym2E
\mathchardef\nshortparallel="3\@ym2F
\mathchardef\nvdash="3\@ym30
\mathchardef\nVdash="3\@ym31
\mathchardef\nvDash="3\@ym32
\mathchardef\nVDash="3\@ym33
\mathchardef\ntrianglerighteq="3\@ym34
\mathchardef\ntrianglelefteq="3\@ym35
\mathchardef\ntriangleleft="3\@ym36
\mathchardef\ntriangleright="3\@ym37
\mathchardef\nleftarrow="3\@ym38
\mathchardef\nrightarrow="3\@ym39
\mathchardef\nLeftarrow="3\@ym3A
\mathchardef\nRightarrow="3\@ym3B
\mathchardef\nLeftrightarrow="3\@ym3C
\mathchardef\nleftrightarrow="3\@ym3D
\mathchardef\divideontimes="2\@ym3E
\mathchardef\varnothing="0\@ym3F
\mathchardef\nexists="0\@ym40
\mathchardef\mho="0\@ym66
\mathchardef\eth="0\@ym67
\mathchardef\eqsim="3\@ym68
\mathchardef\beth="0\@ym69
\mathchardef\gimel="0\@ym6A
\mathchardef\daleth="0\@ym6B
\mathchardef\lessdot="3\@ym6C
\mathchardef\gtrdot="3\@ym6D
\mathchardef\ltimes="2\@ym6E
\mathchardef\rtimes="2\@ym6F
\mathchardef\shortmid="3\@ym70
\mathchardef\shortparallel="3\@ym71
\mathchardef\smallsetminus="2\@ym72
\mathchardef\thicksim="3\@ym73
\mathchardef\thickapprox="3\@ym74
\mathchardef\approxeq="3\@ym75
\mathchardef\succapprox="3\@ym76
\mathchardef\precapprox="3\@ym77
\mathchardef\curvearrowleft="3\@ym78
\mathchardef\curvearrowright="3\@ym79
\mathchardef\digamma="0\@ym7A
\mathchardef\varkappa="0\@ym7B
\mathchardef\hslash="0\@ym7D
\mathchardef\hbar="0\@ym7E
\mathchardef\backepsilon="3\@ym7F


\def\Bbb{\ifmmode\let\next\Bbb@\else
\def\next{\errmessage{Use \string\Bbb\space only in math mode}}\fi\next}
\def\Bbb@#1{{\Bbb@@{#1}}}
\def\Bbb@@#1{\fam\ymfam#1}
\fi


\def\Nulle{0} 
\def\Afe{1}   
\def\Hae{2}   
\def\Hbe{3}   
\def\Hce{4}   
\def\Hde{5}   


\newcount\LastMac       \LastMac=\Nulle

\newskip\half      \half=5.5pt plus 1.5pt minus 2.25pt
\newskip\one       \one=11pt plus 3pt minus 5.5pt
\newskip\onehalf   \onehalf=16.5pt plus 5.5pt minus 8.25pt
\newskip\two       \two=22pt plus 5.5pt minus 11pt

\def\Half{\addvspace{\half}}
\def\One{\addvspace{\one}}
\def\OneHalf{\addvspace{\onehalf}}
\def\Two{\addvspace{\two}}


\def\Raggedright{
  \rightskip=\z@ plus \hsize\relax
}

\def\Fullout{
  \rightskip=\z@\relax
}

\def\Hang#1#2{
  \hangindent=#1%
  \hangafter=#2\relax
}


\newif\ifsp@page
\def\pagestyle#1{\csname ps@#1\endcsname}
\def\thispagestyle#1{\global\sp@pagetrue\gdef\sp@type{#1}}

\def\ps@titlepage{%
  \def\@oddhead{\eightpoint\noindent \the\CatchLine
    \ifprod@font\else\qquad Printed\ \today\fi \hfil}%
  \let\@evenhead=\@oddhead
}

\def\ps@headings{%
  \def\@oddhead{\elevenpoint\it\noindent
    \hfill\the\RightHeader\hskip1.5em\rm\folio}%
  \def\@evenhead{\elevenpoint\noindent
    \folio\hskip1.5em\it\the\LeftHeader\hfill}%
}

\def\ps@plate{%
  \def\@oddhead{\eightpoint\noindent\plt@cap\hfil}%
  \def\@evenhead{\eightpoint\noindent\plt@cap\hfil}%
}



\def\title#1{
  \bgroup
    \vbox to 8pt{\vss}%
    \seventeenpoint
    \Raggedright
    \noindent \strut{\bf #1}\par
  \egroup
}

\def\author#1{
  \bgroup
    \ifnum\LastMac=\Afe \OneHalf\else \vskip 21pt\fi
    \fourteenpoint
    \Raggedright
    \noindent \strut #1\par
    \vskip 3pt%
  \egroup
}

\def\affiliation#1{
  \bgroup
    \vskip -4pt%
    \eightpoint
    \Raggedright
    \noindent \strut {\it #1}\par
  \egroup
  \LastMac=\Afe\relax
}

\def\acceptedline#1{
  \bgroup
    \Two
    \eightpoint
    \Raggedright
    \noindent \strut #1\par
  \egroup
}

\long\def\abstract#1{%
  \bgroup
    \vskip 20pt%
    \everypar{\Hang{11pc}{0}}%
    \noindent{\ninebf ABSTRACT}\par
    \tenpoint
    \Fullout
    \noindent #1\par
  \egroup
}

\long\def\keywords#1{
  \bgroup
    \Half
    \everypar{\Hang{11pc}{0}}%
    \tenpoint
    \Fullout
    \noindent\hbox{\bf Key words:}\ #1\par
  \egroup
}


\def\maketitle{%
  \EndOpening
  \ifsinglecol \else \MakePage\fi
}


\def\pageoffset#1#2{\hoffset=#1\relax\voffset=#2\relax}


\def\Autonumber{
  \global\AutoNumbertrue  
}

\newif\ifAutoNumber \AutoNumberfalse
\newcount\Sec        
\newcount\SecSec
\newcount\SecSecSec

\Sec=\z@

\def\:{\let\@sptoken= } \:  
\def\:{\@xifnch} \expandafter\def\: {\futurelet\@tempc\@ifnch}

\def\@ifnextchar#1#2#3{%
  \let\@tempMACe #1%
  \def\@tempMACa{#2}%
  \def\@tempMACb{#3}%
  \futurelet \@tempMACc\@ifnch%
}

\def\@ifnch{%
\ifx \@tempMACc \@sptoken%
  \let\@tempMACd\@xifnch%
\else%
  \ifx \@tempMACc \@tempMACe%
    \let\@tempMACd\@tempMACa%
  \else%
    \let\@tempMACd\@tempMACb%
  \fi%
\fi%
\@tempMACd%
}

\def\@ifstar#1#2{\@ifnextchar *{\def\@tempMACa*{#1}\@tempMACa}{#2}}

\newskip\@tempskipb

\def\addvspace#1{%
  \ifvmode\else \endgraf\fi%
  \ifdim\lastskip=\z@%
    \vskip #1\relax%
  \else%
    \@tempskipb#1\relax\@xaddvskip%
  \fi%
}

\def\@xaddvskip{%
  \ifdim\lastskip<\@tempskipb%
    \vskip-\lastskip%
    \vskip\@tempskipb\relax%
  \else%
    \ifdim\@tempskipb<\z@%
      \ifdim\lastskip<\z@ \else%
        \advance\@tempskipb\lastskip%
        \vskip-\lastskip\vskip\@tempskipb%
      \fi%
    \fi%
  \fi%
}

\newskip\@tmpSKIP

\def\addpen#1{%
  \ifvmode
    \if@nobreak
    \else
      \ifdim\lastskip=\z@
        \penalty#1\relax
      \else
        \@tmpSKIP=\lastskip
        \vskip -\lastskip
        \penalty#1\vskip\@tmpSKIP
      \fi
    \fi
  \fi
}

\newcount\@clubpen   \@clubpen=\clubpenalty
\newif\if@nobreak    \@nobreakfalse

\def\@noafterindent{%
  \global\@nobreaktrue
  \everypar{\if@nobreak
              \global\@nobreakfalse
              \clubpenalty \@M
              {\setbox\z@\lastbox}%
              \LastMac=\Nulle\relax%
            \else
              \clubpenalty \@clubpen
              \everypar{}%
            \fi}
}

\newcount\gds@cbrk   \gds@cbrk=-300

\def\@nohdbrk{\interlinepenalty \@M\relax}

\let\@par=\par
\def\@restorepar{\def\par{\@par}}

\newif\if@endpe   \@endpefalse
 
\def\@doendpe{\@endpetrue \@nobreakfalse \LastMac=\Nulle\relax%
     \def\par{\@restorepar\everypar{}\par\@endpefalse}%
              \everypar{\setbox\z@\lastbox\everypar{}\@endpefalse}%
}

\def\section{\@ifstar{\@ssection}{\@section}}

\def\@section#1{
  \if@nobreak
    \everypar{}%
    \ifnum\LastMac=\Hae \addvspace{\half}\fi
  \else
    \addpen{\gds@cbrk}%
    \addvspace{\two}%
  \fi
  \bgroup
    \ninepoint\bf
    \Raggedright
    \ifAutoNumber
      \global\advance\Sec \@ne
      \noindent\@nohdbrk\number\Sec\hskip 1pc \uppercase{#1}\par
      \global\SecSec=\z@
    \else
      \noindent\@nohdbrk\uppercase{#1}\par
    \fi
  \egroup
  \nobreak
  \vskip\half
  \nobreak
  \@noafterindent
  \LastMac=\Hae\relax
}

\def\@ssection#1{
  \if@nobreak
    \everypar{}%
    \ifnum\LastMac=\Hae \addvspace{\half}\fi
  \else
    \addpen{\gds@cbrk}%
    \addvspace{\two}%
  \fi
  \bgroup
    \ninepoint\bf
    \Raggedright
    \noindent\@nohdbrk\uppercase{#1}\par
  \egroup
  \nobreak
  \vskip\half
  \nobreak
  \@noafterindent
  \LastMac=\Hae\relax
}

\def\subsection#1{
  \if@nobreak
    \everypar{}%
    \ifnum\LastMac=\Hae \addvspace{1pt plus 1pt minus .5pt}\fi
  \else
    \addpen{\gds@cbrk}%
    \addvspace{\onehalf}%
  \fi
  \bgroup
    \ninepoint\bf
    \Raggedright
    \ifAutoNumber
      \global\advance\SecSec \@ne
      \noindent\@nohdbrk\number\Sec.\number\SecSec \hskip 1pc\relax #1\par
      \global\SecSecSec=\z@
    \else
      \noindent\@nohdbrk #1\par
    \fi
  \egroup
  \nobreak
  \vskip\half
  \nobreak
  \@noafterindent
  \LastMac=\Hbe\relax
}

\def\subsubsection#1{
  \if@nobreak
    \everypar{}%
    \ifnum\LastMac=\Hbe \addvspace{1pt plus 1pt minus .5pt}\fi
  \else
    \addpen{\gds@cbrk}%
    \addvspace{\onehalf}%
  \fi
  \bgroup
    \ninepoint\it
    \Raggedright
    \ifAutoNumber
      \global\advance\SecSecSec \@ne
      \noindent\@nohdbrk\number\Sec.\number\SecSec.\number\SecSecSec
        \hskip 1pc\relax #1\par
    \else
      \noindent\@nohdbrk #1\par
    \fi
  \egroup
  \nobreak
  \vskip\half
  \nobreak
  \@noafterindent
  \LastMac=\Hce\relax
}

\def\paragraph#1{
  \if@nobreak
    \everypar{}%
  \else
    \addpen{\gds@cbrk}%
    \addvspace{\one}%
  \fi%
  \bgroup%
    \ninepoint\it
    \noindent #1\ \nobreak%
  \egroup
  \LastMac=\Hde\relax
  \ignorespaces
}


\let\tx=\relax 


\def\beginlist{%
  \par\if@nobreak \else\addvspace{\half}\fi%
  \bgroup%
    \ninepoint
    \let\item=\list@item%
}

\def\list@item{%
  \par\noindent\hskip 1em\relax%
  \ignorespaces%
}

\def\endlist{\par\egroup\addvspace{\half}\@doendpe}


\def\beginrefs{%
  \par
  \bgroup
    \eightpoint
    \Raggedright
    \let\bibitem=\bib@item
}

\def\bib@item{%
  \par\parindent=1.5em\Hang{1.5em}{1}%
  \everypar={\Hang{1.5em}{1}\ignorespaces}%
  \noindent\ignorespaces
}

\def\endrefs{\par\egroup\@doendpe}


\newtoks\CatchLine

\def\@journal{Mon.\ Not.\ R.\ Astron.\ Soc.\ }  
\def\@pubyear{1994}        
\def\@pagerange{000--000}  
\def\@volume{000}          
\def\@microfiche{}         %

\def\pubyear#1{\gdef\@pubyear{#1}\@makecatchline}
\def\pagerange#1{\gdef\@pagerange{#1}\@makecatchline}
\def\volume#1{\gdef\@volume{#1}\@makecatchline}
\def\microfiche#1{\gdef\@microfiche{and Microfiche\ #1}\@makecatchline}

\def\@makecatchline{%
  \global\CatchLine{%
    {\rm \@journal {\bf \@volume},\ \@pagerange\ (\@pubyear)\ \@microfiche}}%
}

\@makecatchline 

\newtoks\LeftHeader
\def\shortauthor#1{
  \global\LeftHeader{#1}%
}

\newtoks\RightHeader
\def\shorttitle#1{
  \global\RightHeader{#1}%
}

\def\PageHead{
  \begingroup
    \ifsp@page
      \csname ps@\sp@type\endcsname
      \global\sp@pagefalse
    \fi
    \ifodd\pageno
      \let\the@head=\@oddhead
    \else
      \let\the@head=\@evenhead
    \fi
    \vbox to \z@{\vskip-22.5\p@%
      \hbox to \PageWidth{\vbox to8.5\p@{}%
        \the@head
      }%
    \vss}%
  \endgroup
  \nointerlineskip
}

\def\today{%
  \number\day\space
  \ifcase\month\or January\or February\or March\or April\or May\or June\or
    July\or August\or September\or October\or November\or December\fi
  \space\number\year%
}

\def\PageFoot{} 

\def\authorcomment#1{%
  \gdef\PageFoot{%
    \nointerlineskip%
    \vbox to 22pt{\vfil%
      \hbox to \PageWidth{\elevenpoint\noindent \hfil #1 \hfil}}%
  }%
}


\newif\ifplate@page
\newbox\plt@box

\def\beginplatepage{%
  \let\plate=\plate@head
  \let\caption=\fig@caption
  \global\setbox\plt@box=\vbox\bgroup
  \TEMPDIMEN=\PageWidth 
  \hsize=\PageWidth\relax
}

\def\endplatepage{\par\egroup\global\plate@pagetrue}
\def\plate@head#1{\gdef\plt@cap{#1}}


\def\letters{%
  \gdef\folio{\ifnum\pageno<\z@ L\romannumeral-\pageno
    \else L\number\pageno \fi}%
}


\everydisplay{\displaysetup}

\newif\ifeqno
\newif\ifleqno

\def\displaysetup#1$${%
 \displaytest#1\eqno\eqno\displaytest
}

\def\displaytest#1\eqno#2\eqno#3\displaytest{%
 \if!#3!\ldisplaytest#1\leqno\leqno\ldisplaytest
 \else\eqnotrue\leqnofalse\def\eqn{#2}\def\eq{#1}\fi
 \generaldisplay$$}

\def\ldisplaytest#1\leqno#2\leqno#3\ldisplaytest{%
 \def\eq{#1}%
 \if!#3!\eqnofalse\else\eqnotrue\leqnotrue
  \def\eqn{#2}\fi}

\def\generaldisplay{%
\ifeqno \ifleqno 
   \hbox to \hsize{\noindent
     $\displaystyle\eq$\hfil$\displaystyle\eqn$}
  \else
    \hbox to \hsize{\noindent
     $\displaystyle\eq$\hfil$\displaystyle\eqn$}
  \fi
 \else
 \hbox to \hsize{\vbox{\noindent
  $\displaystyle\eq$\hfil}}
 \fi
}


\def\@notice{%
  \par\Two%
  \noindent{\b@ls{11pt}\ninerm This paper has been produced using the
    Blackwell Scientific Publications \TeX\ macros.\par}%
}

\outer\def\bye{\@notice\par\vfill\supereject\end}


\def\start@mess{%
  Monthly notices of the RAS journal style (\@typeface)\space
    v\@version,\space \@verdate.%
}

\everyjob{\Warn{\start@mess}}



\newif\if@debug \@debugfalse  

\def\Print#1{\if@debug\immediate\write16{#1}\else \fi}
\def\Warn#1{\immediate\write16{#1}}
\def\wlog#1{}

\newcount\Iteration 

\def\Single{0} \def\Double{1}                 
\def\Figure{0} \def\Table{1}                  

\def\InStack{0}  
\def\InZoneA{1}
\def\InZoneB{2}
\def\InZoneC{3}

\newcount\TEMPCOUNT 
\newdimen\TEMPDIMEN 
\newbox\TEMPBOX     
\newbox\VOIDBOX     

\newcount\LengthOfStack 
\newcount\MaxItems      
\newcount\StackPointer
\newcount\Point         
\newcount\NextFigure    
\newcount\NextTable     
\newcount\NextItem      

\newcount\StatusStack   
\newcount\NumStack      
\newcount\TypeStack     
\newcount\SpanStack     
\newcount\BoxStack      

\newcount\ItemSTATUS    
\newcount\ItemNUMBER    
\newcount\ItemTYPE      
\newcount\ItemSPAN      
\newbox\ItemBOX         
\newdimen\ItemSIZE      

\newdimen\PageHeight    
\newdimen\TextLeading   
\newdimen\Feathering    
\newcount\LinesPerPage  
\newdimen\ColumnWidth   
\newdimen\ColumnGap     
\newdimen\PageWidth     
\newdimen\BodgeHeight   
\newcount\Leading       

\newdimen\ZoneBSize  
\newdimen\TextSize   
\newbox\ZoneABOX     
\newbox\ZoneBBOX     
\newbox\ZoneCBOX     

\newif\ifFirstSingleItem
\newif\ifFirstZoneA
\newif\ifMakePageInComplete
\newif\ifMoreFigures \MoreFiguresfalse 
\newif\ifMoreTables  \MoreTablesfalse  

\newif\ifFigInZoneB 
\newif\ifFigInZoneC 
\newif\ifTabInZoneB 
\newif\ifTabInZoneC

\newif\ifZoneAFullPage

\newbox\MidBOX    
\newbox\LeftBOX
\newbox\RightBOX
\newbox\PageBOX   

\newif\ifLeftCOL  
\LeftCOLtrue

\newdimen\ZoneBAdjust

\newcount\ItemFits
\def\Yes{1}
\def\No{2}


\MaxItems=15
\NextFigure=\z@        
\NextTable=\@ne

\BodgeHeight=6pt
\TextLeading=11pt    
\Leading=11
\Feathering=\z@      
\LinesPerPage=61     
\topskip=\TextLeading
\ColumnWidth=20pc    
\ColumnGap=2pc       

\newskip\ItemSepamount  
\ItemSepamount=\TextLeading plus \TextLeading minus 4pt

\parskip=\z@ plus .1pt
\parindent=18pt
\widowpenalty=\z@
\clubpenalty=10000
\tolerance=1500
\hbadness=1500
\abovedisplayskip=6pt plus 2pt minus 2pt
\belowdisplayskip=6pt plus 2pt minus 2pt
\abovedisplayshortskip=6pt plus 2pt minus 2pt
\belowdisplayshortskip=6pt plus 2pt minus 2pt

\ninepoint 


\PageHeight=682pt

\PageWidth=2\ColumnWidth
\advance\PageWidth by \ColumnGap

\pagestyle{headings}




\newcount\DUMMY \StatusStack=\allocationnumber
\newcount\DUMMY \newcount\DUMMY \newcount\DUMMY 
\newcount\DUMMY \newcount\DUMMY \newcount\DUMMY 
\newcount\DUMMY \newcount\DUMMY \newcount\DUMMY
\newcount\DUMMY \newcount\DUMMY \newcount\DUMMY 
\newcount\DUMMY \newcount\DUMMY \newcount\DUMMY

\newcount\DUMMY \NumStack=\allocationnumber
\newcount\DUMMY \newcount\DUMMY \newcount\DUMMY 
\newcount\DUMMY \newcount\DUMMY \newcount\DUMMY 
\newcount\DUMMY \newcount\DUMMY \newcount\DUMMY 
\newcount\DUMMY \newcount\DUMMY \newcount\DUMMY 
\newcount\DUMMY \newcount\DUMMY \newcount\DUMMY

\newcount\DUMMY \TypeStack=\allocationnumber
\newcount\DUMMY \newcount\DUMMY \newcount\DUMMY 
\newcount\DUMMY \newcount\DUMMY \newcount\DUMMY 
\newcount\DUMMY \newcount\DUMMY \newcount\DUMMY 
\newcount\DUMMY \newcount\DUMMY \newcount\DUMMY 
\newcount\DUMMY \newcount\DUMMY \newcount\DUMMY

\newcount\DUMMY \SpanStack=\allocationnumber
\newcount\DUMMY \newcount\DUMMY \newcount\DUMMY 
\newcount\DUMMY \newcount\DUMMY \newcount\DUMMY 
\newcount\DUMMY \newcount\DUMMY \newcount\DUMMY 
\newcount\DUMMY \newcount\DUMMY \newcount\DUMMY 
\newcount\DUMMY \newcount\DUMMY \newcount\DUMMY

\newbox\DUMMY   \BoxStack=\allocationnumber
\newbox\DUMMY   \newbox\DUMMY \newbox\DUMMY 
\newbox\DUMMY   \newbox\DUMMY \newbox\DUMMY 
\newbox\DUMMY   \newbox\DUMMY \newbox\DUMMY 
\newbox\DUMMY   \newbox\DUMMY \newbox\DUMMY 
\newbox\DUMMY   \newbox\DUMMY \newbox\DUMMY

\def\wlog{\immediate\write\m@ne}


\def\GetItemAll#1{%
 \GetItemSTATUS{#1}
 \GetItemNUMBER{#1}
 \GetItemTYPE{#1}
 \GetItemSPAN{#1}
 \GetItemBOX{#1}
}

\def\GetItemSTATUS#1{%
 \Point=\StatusStack
 \advance\Point by #1
 \global\ItemSTATUS=\count\Point
}

\def\GetItemNUMBER#1{%
 \Point=\NumStack
 \advance\Point by #1
 \global\ItemNUMBER=\count\Point
}

\def\GetItemTYPE#1{%
 \Point=\TypeStack
 \advance\Point by #1
 \global\ItemTYPE=\count\Point
}

\def\GetItemSPAN#1{%
 \Point\SpanStack
 \advance\Point by #1
 \global\ItemSPAN=\count\Point
}

\def\GetItemBOX#1{%
 \Point=\BoxStack
 \advance\Point by #1
 \global\setbox\ItemBOX=\vbox{\copy\Point}
 \global\ItemSIZE=\ht\ItemBOX
 \global\advance\ItemSIZE by \dp\ItemBOX
 \TEMPCOUNT=\ItemSIZE
 \divide\TEMPCOUNT by \Leading
 \divide\TEMPCOUNT by 65536
 \advance\TEMPCOUNT \@ne
 \ItemSIZE=\TEMPCOUNT pt
 \global\multiply\ItemSIZE by \Leading
}


\def\JoinStack{%
 \ifnum\LengthOfStack=\MaxItems 
  \Warn{WARNING: Stack is full...some items will be lost!}
 \else
  \Point=\StatusStack
  \advance\Point by \LengthOfStack
  \global\count\Point=\ItemSTATUS
  \Point=\NumStack
  \advance\Point by \LengthOfStack
  \global\count\Point=\ItemNUMBER
  \Point=\TypeStack
  \advance\Point by \LengthOfStack
  \global\count\Point=\ItemTYPE
  \Point\SpanStack
  \advance\Point by \LengthOfStack
  \global\count\Point=\ItemSPAN
  \Point=\BoxStack
  \advance\Point by \LengthOfStack
  \global\setbox\Point=\vbox{\copy\ItemBOX}
  \global\advance\LengthOfStack \@ne
  \ifnum\ItemTYPE=\Figure 
   \global\MoreFigurestrue
  \else
   \global\MoreTablestrue
  \fi
 \fi
}


\def\LeaveStack#1{%
 {\Iteration=#1
 \loop
 \ifnum\Iteration<\LengthOfStack
  \advance\Iteration \@ne
  \GetItemSTATUS{\Iteration}
   \advance\Point by \m@ne
   \global\count\Point=\ItemSTATUS
  \GetItemNUMBER{\Iteration}
   \advance\Point by \m@ne
   \global\count\Point=\ItemNUMBER
  \GetItemTYPE{\Iteration}
   \advance\Point by \m@ne
   \global\count\Point=\ItemTYPE
  \GetItemSPAN{\Iteration}
   \advance\Point by \m@ne
   \global\count\Point=\ItemSPAN
  \GetItemBOX{\Iteration}
   \advance\Point by \m@ne
   \global\setbox\Point=\vbox{\copy\ItemBOX}
 \repeat}
 \global\advance\LengthOfStack by \m@ne
}


\newif\ifStackNotClean

\def\CleanStack{%
 \StackNotCleantrue
 {\Iteration=\z@
  \loop
   \ifStackNotClean
    \GetItemSTATUS{\Iteration}
    \ifnum\ItemSTATUS=\InStack
     \advance\Iteration \@ne
     \else
      \LeaveStack{\Iteration}
    \fi
   \ifnum\LengthOfStack<\Iteration
    \StackNotCleanfalse
   \fi
 \repeat}
}


\def\FindItem#1#2{%
 \global\StackPointer=\m@ne 
 {\Iteration=\z@
  \loop
  \ifnum\Iteration<\LengthOfStack
   \GetItemSTATUS{\Iteration}
   \ifnum\ItemSTATUS=\InStack
    \GetItemTYPE{\Iteration}
    \ifnum\ItemTYPE=#1
     \GetItemNUMBER{\Iteration}
     \ifnum\ItemNUMBER=#2
      \global\StackPointer=\Iteration
      \Iteration=\LengthOfStack 
     \fi
    \fi
   \fi
  \advance\Iteration \@ne
 \repeat}
}


\def\FindNext{%
 \global\StackPointer=\m@ne 
 {\Iteration=\z@
  \loop
  \ifnum\Iteration<\LengthOfStack
   \GetItemSTATUS{\Iteration}
   \ifnum\ItemSTATUS=\InStack
    \GetItemTYPE{\Iteration}
   \ifnum\ItemTYPE=\Figure
    \ifMoreFigures
      \global\NextItem=\Figure
      \global\StackPointer=\Iteration
      \Iteration=\LengthOfStack 
    \fi
   \fi
   \ifnum\ItemTYPE=\Table
    \ifMoreTables
      \global\NextItem=\Table
      \global\StackPointer=\Iteration
      \Iteration=\LengthOfStack 
    \fi
   \fi
  \fi
  \advance\Iteration \@ne
 \repeat}
}


\def\ChangeStatus#1#2{%
 \Point=\StatusStack
 \advance\Point by #1
 \global\count\Point=#2
}



\def\Zone{\InZoneA}

\ZoneBAdjust=\z@

\def\MakePage{
 \global\ZoneBSize=\PageHeight
 \global\TextSize=\ZoneBSize
 \global\ZoneAFullPagefalse
 \global\topskip=\TextLeading
 \MakePageInCompletetrue
 \MoreFigurestrue
 \MoreTablestrue
 \FigInZoneBfalse
 \FigInZoneCfalse
 \TabInZoneBfalse
 \TabInZoneCfalse
 \global\FirstSingleItemtrue
 \global\FirstZoneAtrue
 \global\setbox\ZoneABOX=\box\VOIDBOX
 \global\setbox\ZoneBBOX=\box\VOIDBOX
 \global\setbox\ZoneCBOX=\box\VOIDBOX
 \loop
  \ifMakePageInComplete
 \FindNext
 \ifnum\StackPointer=\m@ne
  \NextItem=\m@ne
  \MoreFiguresfalse
  \MoreTablesfalse
 \fi
 \ifnum\NextItem=\Figure
   \FindItem{\Figure}{\NextFigure}
   \ifnum\StackPointer=\m@ne \global\MoreFiguresfalse
   \else
    \GetItemSPAN{\StackPointer}
    \ifnum\ItemSPAN=\Single \def\Zone{\InZoneB}\relax
     \ifFigInZoneC \global\MoreFiguresfalse\fi
    \else
     \def\Zone{\InZoneA}
     \ifFigInZoneB \def\Zone{\InZoneC}\fi
    \fi
   \fi
   \ifMoreFigures\Print{}\FigureItems\fi
 \fi
\ifnum\NextItem=\Table
   \FindItem{\Table}{\NextTable}
   \ifnum\StackPointer=\m@ne \global\MoreTablesfalse
   \else
    \GetItemSPAN{\StackPointer}
    \ifnum\ItemSPAN=\Single\relax
     \ifTabInZoneC \global\MoreTablesfalse\fi
    \else
     \def\Zone{\InZoneA}
     \ifTabInZoneB \def\Zone{\InZoneC}\fi
    \fi
   \fi
   \ifMoreTables\Print{}\TableItems\fi
 \fi
   \MakePageInCompletefalse 
   \ifMoreFigures\MakePageInCompletetrue\fi
   \ifMoreTables\MakePageInCompletetrue\fi
 \repeat
 \ifZoneAFullPage
  \global\TextSize=\z@
  \global\ZoneBSize=\z@
  \global\vsize=\z@\relax
  \global\topskip=\z@\relax
  \vbox to \z@{\vss}
  \eject
 \else
 \global\advance\ZoneBSize by -\ZoneBAdjust
 \global\vsize=\ZoneBSize
 \global\hsize=\ColumnWidth
 \global\ZoneBAdjust=\z@
 \ifdim\TextSize<23pt
 \Warn{}
 \Warn{* Making column fall short: TextSize=\the\TextSize *}
 \vskip-\lastskip\eject\fi
 \fi
}

\def\MakeRightCol{
 \global\TextSize=\ZoneBSize
 \MakePageInCompletetrue
 \MoreFigurestrue
 \MoreTablestrue
 \global\FirstSingleItemtrue
 \global\setbox\ZoneBBOX=\box\VOIDBOX
 \def\Zone{\InZoneB}
 \loop
  \ifMakePageInComplete
 \FindNext
 \ifnum\StackPointer=\m@ne
  \NextItem=\m@ne
  \MoreFiguresfalse
  \MoreTablesfalse
 \fi
 \ifnum\NextItem=\Figure
   \FindItem{\Figure}{\NextFigure}
   \ifnum\StackPointer=\m@ne \MoreFiguresfalse
   \else
    \GetItemSPAN{\StackPointer}
    \ifnum\ItemSPAN=\Double\relax
     \MoreFiguresfalse\fi
   \fi
   \ifMoreFigures\Print{}\FigureItems\fi
 \fi
 \ifnum\NextItem=\Table
   \FindItem{\Table}{\NextTable}
   \ifnum\StackPointer=\m@ne \MoreTablesfalse
   \else
    \GetItemSPAN{\StackPointer}
    \ifnum\ItemSPAN=\Double\relax
     \MoreTablesfalse\fi
   \fi
   \ifMoreTables\Print{}\TableItems\fi
 \fi
   \MakePageInCompletefalse 
   \ifMoreFigures\MakePageInCompletetrue\fi
   \ifMoreTables\MakePageInCompletetrue\fi
 \repeat
 \ifZoneAFullPage
  \global\TextSize=\z@
  \global\ZoneBSize=\z@
  \global\vsize=\z@\relax
  \global\topskip=\z@\relax
  \vbox to \z@{\vss}
  \eject
 \else
 \global\vsize=\ZoneBSize
 \global\hsize=\ColumnWidth
 \ifdim\TextSize<23pt
 \Warn{}
 \Warn{* Making column fall short: TextSize=\the\TextSize *}
 \vskip-\lastskip\eject\fi
\fi
}

\def\FigureItems{
 \Print{Considering...}
 \ShowItem{\StackPointer}
 \GetItemBOX{\StackPointer} 
 \GetItemSPAN{\StackPointer}
  \CheckFitInZone 
  \ifnum\ItemFits=\Yes
   \ifnum\ItemSPAN=\Single
     \ChangeStatus{\StackPointer}{\InZoneB} 
     \global\FigInZoneBtrue
     \ifFirstSingleItem
      \hbox{}\vskip-\BodgeHeight
     \global\advance\ItemSIZE by \TextLeading
     \fi
     \unvbox\ItemBOX\ItemSep
     \global\FirstSingleItemfalse
     \global\advance\TextSize by -\ItemSIZE
     \global\advance\TextSize by -\TextLeading
   \else
    \ifFirstZoneA
     \global\advance\ItemSIZE by \TextLeading
     \global\FirstZoneAfalse\fi
    \global\advance\TextSize by -\ItemSIZE
    \global\advance\TextSize by -\TextLeading
    \global\advance\ZoneBSize by -\ItemSIZE
    \global\advance\ZoneBSize by -\TextLeading
    \ifFigInZoneB\relax
     \else
     \ifdim\TextSize<3\TextLeading
     \global\ZoneAFullPagetrue
     \fi
    \fi
    \ChangeStatus{\StackPointer}{\Zone}
    \ifnum\Zone=\InZoneC \global\FigInZoneCtrue\fi
  \fi
   \Print{TextSize=\the\TextSize}
   \Print{ZoneBSize=\the\ZoneBSize}
  \global\advance\NextFigure \@ne
   \Print{This figure has been placed.}
  \else
   \Print{No space available for this figure...holding over.}
   \Print{}
   \global\MoreFiguresfalse
  \fi
}

\def\TableItems{
 \Print{Considering...}
 \ShowItem{\StackPointer}
 \GetItemBOX{\StackPointer} 
 \GetItemSPAN{\StackPointer}
  \CheckFitInZone 
  \ifnum\ItemFits=\Yes
   \ifnum\ItemSPAN=\Single
    \ChangeStatus{\StackPointer}{\InZoneB}
     \global\TabInZoneBtrue
     \ifFirstSingleItem
      \hbox{}\vskip-\BodgeHeight
     \global\advance\ItemSIZE by \TextLeading
     \fi
     \unvbox\ItemBOX\ItemSep
     \global\FirstSingleItemfalse
     \global\advance\TextSize by -\ItemSIZE
     \global\advance\TextSize by -\TextLeading
   \else
    \ifFirstZoneA
    \global\advance\ItemSIZE by \TextLeading
    \global\FirstZoneAfalse\fi
    \global\advance\TextSize by -\ItemSIZE
    \global\advance\TextSize by -\TextLeading
    \global\advance\ZoneBSize by -\ItemSIZE
    \global\advance\ZoneBSize by -\TextLeading
    \ifFigInZoneB\relax
     \else
     \ifdim\TextSize<3\TextLeading
     \global\ZoneAFullPagetrue
     \fi
    \fi
    \ChangeStatus{\StackPointer}{\Zone}
    \ifnum\Zone=\InZoneC \global\TabInZoneCtrue\fi
   \fi
  \global\advance\NextTable \@ne
   \Print{This table has been placed.}
  \else
  \Print{No space available for this table...holding over.}
   \Print{}
   \global\MoreTablesfalse
  \fi
}


\def\CheckFitInZone{%
{\advance\TextSize by -\ItemSIZE
 \advance\TextSize by -\TextLeading
 \ifFirstSingleItem
  \advance\TextSize by \TextLeading
 \fi
 \ifnum\Zone=\InZoneA\relax
  \else \advance\TextSize by -\ZoneBAdjust
 \fi
 \ifdim\TextSize<3\TextLeading \global\ItemFits=\No
 \else \global\ItemFits=\Yes\fi}
}

\def\BeginOpening{%
  \thispagestyle{titlepage}%
  \global\setbox\ItemBOX=\vbox\bgroup%
    \hsize=\PageWidth%
    \hrule height \z@
    \ifsinglecol\vskip 6pt\fi 
}

\let\begintopmatter=\BeginOpening  

\def\EndOpening{%
  \One
  \egroup
  \ifsinglecol
    \box\ItemBOX%
    \vskip\TextLeading plus 2\TextLeading
    \@noafterindent
  \else
    \ItemNUMBER=\z@%
    \ItemTYPE=\Figure
    \ItemSPAN=\Double
    \ItemSTATUS=\InStack
    \JoinStack
  \fi
}


\newif\if@here  \@herefalse

\def\no@float{\global\@heretrue}
\let\nofloat=\relax 

\def\beginfigure{%
  \@ifstar{\global\@dfloattrue \@bfigure}{\global\@dfloatfalse \@bfigure}%
}

\def\@bfigure#1{%
  \par
  \if@dfloat
    \ItemSPAN=\Double
    \TEMPDIMEN=\PageWidth
  \else
    \ItemSPAN=\Single
    \TEMPDIMEN=\ColumnWidth
  \fi
  \ifsinglecol
    \TEMPDIMEN=\PageWidth
  \else
    \ItemSTATUS=\InStack
    \ItemNUMBER=#1%
    \ItemTYPE=\Figure
  \fi
  \bgroup
    \hsize=\TEMPDIMEN
    \global\setbox\ItemBOX=\vbox\bgroup
      \eightpoint\nostb@ls{10pt}%
      \let\caption=\fig@caption
      \ifsinglecol \let\nofloat=\no@float\fi
}

\def\fig@caption#1{%
  \vskip 5.5pt plus 6pt%
  \bgroup 
    \eightpoint\nostb@ls{10pt}%
    \setbox\TEMPBOX=\hbox{#1}%
    \ifdim\wd\TEMPBOX>\TEMPDIMEN
      \noindent \unhbox\TEMPBOX\par
    \else
      \hbox to \hsize{\hfil\unhbox\TEMPBOX\hfil}%
    \fi
  \egroup
}

\def\endfigure{%
  \par\egroup 
  \egroup
  \ifsinglecol
    \if@here \midinsert\global\@herefalse\else \topinsert\fi
      \unvbox\ItemBOX
    \endinsert
  \else
    \JoinStack
    \Print{Processing source for figure \the\ItemNUMBER}%
  \fi
}


\newbox\tab@cap@box
\def\tab@caption#1{\global\setbox\tab@cap@box=\hbox{#1\par}}

\newtoks\tab@txt@toks
\long\def\tab@txt#1{\global\tab@txt@toks={#1}\global\table@txttrue}

\newif\iftable@txt  \table@txtfalse
\newif\if@dfloat    \@dfloatfalse

\def\begintable{%
  \@ifstar{\global\@dfloattrue \@btable}{\global\@dfloatfalse \@btable}%
}

\def\@btable#1{%
  \par
  \if@dfloat
    \ItemSPAN=\Double
    \TEMPDIMEN=\PageWidth
  \else
    \ItemSPAN=\Single
    \TEMPDIMEN=\ColumnWidth
  \fi
  \ifsinglecol
    \TEMPDIMEN=\PageWidth
  \else
    \ItemSTATUS=\InStack
    \ItemNUMBER=#1%
    \ItemTYPE=\Table
  \fi
  \bgroup
    \eightpoint\nostb@ls{10pt}%
    \global\setbox\ItemBOX=\vbox\bgroup
      \let\caption=\tab@caption
      \let\tabletext=\tab@txt
      \ifsinglecol \let\nofloat=\no@float\fi
}

\def\endtable{%
  \par\egroup 
  \egroup
  \setbox\TEMPBOX=\hbox to \TEMPDIMEN{%
    \hss
    \vbox{%
      \hsize=\wd\ItemBOX
      \ifvoid\tab@cap@box
      \else
        \noindent\unhbox\tab@cap@box
        \vskip 5.5pt plus 6pt%
      \fi
      \box\ItemBOX
      \iftable@txt
        \vskip 10pt%
        \eightpoint\nostb@ls{10pt}%
        \noindent\the\tab@txt@toks
        \global\table@txtfalse
      \fi
    }%
    \hss
  }%
  \ifsinglecol
    \if@here \midinsert\global\@herefalse\else \topinsert\fi
      \box\TEMPBOX
    \endinsert
  \else
    \global\setbox\ItemBOX=\box\TEMPBOX
    \JoinStack
    \Print{Processing source for table \the\ItemNUMBER}%
  \fi
}

\def\UnloadZoneA{%
\FirstZoneAtrue
 \Iteration=\z@
  \loop
   \ifnum\Iteration<\LengthOfStack
    \GetItemSTATUS{\Iteration}
    \ifnum\ItemSTATUS=\InZoneA
     \GetItemBOX{\Iteration}
     \ifFirstZoneA \vbox to \BodgeHeight{\vfil}%
     \FirstZoneAfalse\fi
     \unvbox\ItemBOX\ItemSep
     \LeaveStack{\Iteration}
     \else
     \advance\Iteration \@ne
   \fi
 \repeat
}

\def\UnloadZoneC{%
\Iteration=\z@
  \loop
   \ifnum\Iteration<\LengthOfStack
    \GetItemSTATUS{\Iteration}
    \ifnum\ItemSTATUS=\InZoneC
     \GetItemBOX{\Iteration}
     \ItemSep\unvbox\ItemBOX
     \LeaveStack{\Iteration}
     \else
     \advance\Iteration \@ne
   \fi
 \repeat
}


\def\ShowItem#1{
  {\GetItemAll{#1}
  \Print{\the#1:
  {TYPE=\ifnum\ItemTYPE=\Figure Figure\else Table\fi}
  {NUMBER=\the\ItemNUMBER}
  {SPAN=\ifnum\ItemSPAN=\Single Single\else Double\fi}
  {SIZE=\the\ItemSIZE}}}
}

\def\ShowStack{%
 \Print{}
 \Print{LengthOfStack = \the\LengthOfStack}
 \ifnum\LengthOfStack=\z@ \Print{Stack is empty}\fi
 \Iteration=\z@
 \loop
 \ifnum\Iteration<\LengthOfStack
  \ShowItem{\Iteration}
  \advance\Iteration \@ne
 \repeat
}

\def\B#1#2{%
\hbox{\vrule\kern-0.4pt\vbox to #2{%
\hrule width #1\vfill\hrule}\kern-0.4pt\vrule}
}


\newif\ifsinglecol   \singlecolfalse

\def\onecolumn{%
  \global\output={\singlecoloutput}%
  \global\hsize=\PageWidth
  \global\vsize=\PageHeight
  \global\ColumnWidth=\hsize
  \global\TextLeading=12pt
  \global\Leading=12
  \global\singlecoltrue
  \global\let\onecolumn=\relax
  \global\let\footnote=\sing@footnote
  \global\let\vfootnote=\sing@vfootnote
  \ninepoint 
  \message{(Single column)}%
}

\def\singlecoloutput{%
  \shipout\vbox{\PageHead\pagebody\PageFoot}%
  \advancepageno
  \ifplate@page
    \shipout\vbox{%
      \sp@pagetrue
      \def\sp@type{plate}%
      \global\plate@pagefalse
      \PageHead\vbox to \PageHeight{\unvbox\plt@box\vfil}\PageFoot%
    }%
    \message{[plate]}%
    \advancepageno
  \fi
  \ifnum\outputpenalty>-\@MM \else\dosupereject\fi%
}

\def\ItemSep{\vskip\ItemSepamount\relax}

\def\ItemSepbreak{\par\ifdim\lastskip<\ItemSepamount
  \removelastskip\penalty-200\ItemSep\fi%
}


\let\@@endinsert=\endinsert 

\def\endinsert{\egroup 
  \if@mid \dimen@\ht\z@ \advance\dimen@\dp\z@ \advance\dimen@12\p@
    \advance\dimen@\pagetotal \advance\dimen@-\pageshrink
    \ifdim\dimen@>\pagegoal\@midfalse\p@gefalse\fi\fi
  \if@mid \ItemSep\box\z@\ItemSepbreak
  \else\insert\topins{\penalty100 
    \splittopskip\z@skip
    \splitmaxdepth\maxdimen \floatingpenalty\z@
    \ifp@ge \dimen@\dp\z@
    \vbox to\vsize{\unvbox\z@\kern-\dimen@}
    \else \box\z@\nobreak\ItemSep\fi}\fi\endgroup%
}


\def\gobbleone#1{}
\def\gobbletwo#1#2{}
\let\footnote=\gobbletwo 
\let\vfootnote=\gobbleone

\def\sing@footnote#1{\let\@sf\empty 
  \ifhmode\edef\@sf{\spacefactor\the\spacefactor}\/\fi
  \hbox{$^{\hbox{\eightpoint #1}}$}\@sf\sing@vfootnote{#1}%
}

\def\sing@vfootnote#1{\insert\footins\bgroup\eightpoint\b@ls{9pt}%
  \interlinepenalty\interfootnotelinepenalty
  \splittopskip\ht\strutbox 
  \splitmaxdepth\dp\strutbox \floatingpenalty\@MM
  \leftskip\z@skip \rightskip\z@skip \spaceskip\z@skip \xspaceskip\z@skip
  \noindent $^{\scriptstyle\hbox{#1}}$\hskip 4pt%
    \footstrut\futurelet\next\fo@t%
}

\def\footnoterule{\kern-3\p@ \hrule height \z@ \kern 3\p@}

\skip\footins=19.5pt plus 12pt minus 1pt
\count\footins=1000
\dimen\footins=\maxdimen


\def\landscape{%
  \global\TEMPDIMEN=\PageWidth
  \global\PageWidth=\PageHeight
  \global\PageHeight=\TEMPDIMEN
  \global\let\landscape=\relax
  \onecolumn
  \message{(landscape)}%
  \raggedbottom
}


\output{%
  \ifLeftCOL
    \global\setbox\LeftBOX=\vbox to \ZoneBSize{\box255\unvbox\ZoneBBOX}%
    \global\LeftCOLfalse
    \MakeRightCol
  \else
    \setbox\RightBOX=\vbox to \ZoneBSize{\box255\unvbox\ZoneBBOX}%
    \setbox\MidBOX=\hbox{\box\LeftBOX\hskip\ColumnGap\box\RightBOX}%
    \setbox\PageBOX=\vbox to \PageHeight{%
      \UnloadZoneA\box\MidBOX\UnloadZoneC}%
    \shipout\vbox{\PageHead\box\PageBOX\PageFoot}%
    \advancepageno
    \ifplate@page
      \shipout\vbox{%
        \sp@pagetrue
        \def\sp@type{plate}%
        \global\plate@pagefalse
        \PageHead\vbox to \PageHeight{\unvbox\plt@box\vfil}\PageFoot%
      }%
      \message{[plate]}%
      \advancepageno
    \fi
    \global\LeftCOLtrue
    \CleanStack
    \MakePage
  \fi
}


\Warn{\start@mess}


\catcode `\@=12 




\let\sec=\section
\let\ssec=\subsection


\def\bigstrut{\vrule width0pt height0.6truecm}
\font\japit = cmti10 at 11truept
\def\ss{\scriptscriptstyle\rm}
\def\ref{\parskip =0pt\par\noindent\hangindent\parindent
    \parskip =2ex plus .5ex minus .1ex}
\def\sqamin{\hbox{arcmin$^2$}} 
\def\gs{\mathrel{\lower0.6ex\hbox{$\buildrel {\textstyle >}
 \over {\scriptstyle \sim}$}}}
\def\ls{\mathrel{\lower0.6ex\hbox{$\buildrel {\textstyle <}
 \over {\scriptstyle \sim}$}}}
\newcount\equationo
\equationo = 0
\def\leftdisplay#1$${\leftline{$\displaystyle{#1}$
  \global\advance\equationo by1\hfill (\the\equationo )}$$}
\everydisplay{\leftdisplay}

\font\bbbit=cmbxti10 at 17truept

\def\etal{{\rm et~al.}}

\def\Msun{\hbox{$M_{\odot}$}}

\def\name#1{{\it #1\/}}
\def\vol#1{{\bf #1\/}}

\def\MNRAS{{\jfont Mon. Not. R. Astr. Soc.}}
\def\ApJ{{\jfont Astrophys. J.}}
\def\AstrAst{{\jfont Astr. Astrophys.}}
\def\AstrAstSup{{\jfont Astr. Astrophys. Suppl. Ser.}}
\def\AstrJ{{\jfont Astr. J.}}

%

\pageoffset{-0.8cm}{0.2cm}

\Autonumber  



\begintopmatter  

\vglue-1.9truecm
\centerline{\japit Accepted for publication in Monthly Notices of the R.A.S.}
\vglue 2.0truecm

\title{An imaging {\bbbit K\/}-band survey -- II: The redshift survey
and galaxy evolution in the infrared}

\author{Karl Glazebrook$^{1,3}$,
J.A. Peacock$^2$, L. Miller$^2$ and C.A. Collins$^{2,4}$ }

\affiliation{$^1$Institute for Astronomy, \bigstrut University of Edinburgh,
Blackford Hill, Edinburgh EH9 3HJ, UK \hfill\break
$^2$Royal Observatory, Blackford Hill, Edinburgh EH9 3HJ, UK\hfill\break
$^3$Present address: Institute of Astronomy, Madingley Road,
Cambridge CB3 0HA, UK\hfill\break
$^4$Present address: Chemical \& Physical Sciences, Liverpool John Moores 
University, Byrom Street, Liverpool L3 3AF, UK}

\shortauthor{K. Glazebrook, J.A. Peacock, L. Miller and C.A. Collins}

\shorttitle{Imaging $K$-band survey -- II.}


\abstract {We present further results from an imaging $K$-band survey
of $552\,\sqamin$, complete to a 5$\sigma$ limit
of $K\simeq 17.3$.  This paper describes a redshift survey
of 124 galaxies, and addresses the colours of faint galaxies and the
evolution of the $K$-band luminosity function.  The optical-to-infrared
colours are consistent with the range expected from synthetic galaxy
spectra, although there are some cases of very red nuclei. These may 
possibly be attributed to either extinction or metallicity gradients.  Our
data show no evidence for evolution of the $K$-band luminosity function
at $z<0.5$, and the results are well described by a Schechter function
with $M_K^*=-22.75\pm0.13+5\log_{10}h$ and 
$\phi^*=0.026\pm0.003\, h^3\, {\rm Mpc^{-3}}$. 
This is a somewhat higher normalization than has
been found by previous workers, and it removes much of the excess in
faint $K$ and $b_{\ss J}$ counts with respect to a no-evolution model.
However, we do find evidence for evolution at $z>0.5$: $M_K^*$ is
approximately 0.75 mag. brighter at $z=1$.
This luminosity evolution is balanced by a
reduced normalization at high redshift: the total luminosity density
is required to be approximately constant in order not to
exceed the faint counts. 
The overall evolution is thus {\it opposite\/} to that expected
in simple merger-dominated models; we briefly consider possible
interpretations of this result.
} 

\maketitle  

\sec{INTRODUCTION}

The development of two-dimensional near-infrared detectors has finally
made it possible to survey substantial areas of the 
sky at these wavelengths to cosmologically interesting depths. Two
surveys covering more than several hundred
\sqamin\ of the sky have recently been completed: 
the Hawaii Surveys described by Gardner \etal\ (1993),
Cowie \etal\ (1994) and Songaila \etal\  (1994), and the
Edinburgh Survey described by
Glazebrook \etal\ (1994 -- hereafter referred to as Paper I).

This is the second paper concerning our $K$-band redshift
survey covering $552\,\sqamin$.  Paper I discussed in detail the
construction and calibration of this survey, and the associated optical
CCD imaging for all the fields. It also presented our results for the
$K$-band star and galaxy counts. This paper is concerned
with a $K$-selected redshift survey and the 
resulting colour-redshift and luminosity function analyses.

A study of galaxy evolution in the near infrared is
of great interest. Historically the main evidence for bulk evolution in
samples of field galaxies derives from optical surveys, culminating
in the `faint blue galaxy problem': number-magnitude counts over
$15<b_{\ss J}<28$ are much steeper than predicted by a non-evolving
model (see Ellis 1990 for a review).
Optically-selected redshift surveys have shown this faint excess
to be a population of very blue objects,
evolved mainly in density rather than blue luminosity
(Broadhurst \etal\ 1988; Colless \etal\ 1991 and Glazebrook \etal\ 
1995a). Various models have been proposed to explain this
observation: 
Koo \etal\  (1993) have proposed radical alteration of the
local luminosity function by introducing a large dwarf galaxy
component. Similarly, Cowie \etal\  (1991) and Babul
\& Rees (1992) have proposed a model in which the blue population
consists of a new population of dwarf galaxies undergoing an initial
starburst at $z\sim 0.4$ and fading to invisibility by the
present day.  A different model has been proposed by Broadhurst, Ellis
\& Glazebrook (1992: BEG) and Rocca-Volmerange \& Guiderdoni (1990) in which
a large amount of galaxy-galaxy merging has occured in the field
population in the recent past,
as might be expected in CDM-like theories in which structure grows by
hierarchical growth (e.g. Carlberg 1992).
However, the most recent possible explanation is the simplest: that bright
surveys are not sufficiently sensitive to galaxies of
low surface brightness (McGaugh 1994).
This suggestion of incompleteness in the bright counts
accords with the work of Metcalfe \etal\  (1991 \& 1994), who
have argued that the fainter data for $17<b_{\ss J}<22$ can
then be explained by a high local normalisation to the luminosity
function.

In view of these controversies, an independent approach to galaxy
evolution is clearly attractive, and
the $K$ band is in many ways preferable to the optical.
The optical work samples the rest-frame UV at $z\gs 0.3$, so that
the optical luminosity depends sensitively on the rate of
star formation. Also, most of the
optical light in galaxies comes from massive OB stars which are only a
small fraction of the total stellar mass of the galaxy. Thus a dwarf
galaxy undergoing a powerful starburst can attain a $b_{\ss J}$
luminosity identical to that of a giant spiral 
or elliptical galaxy evolving quiescently.
Since galaxies at high redshift are unresolved from the ground,
these very dissimilar systems can be indistinguishable
in faint optical imaging data.

In contrast, the near-infrared light in galaxies 
is produced by giants drawn from the population of
old evolved stars which dominate the stellar mass.  
Moreover the K-correction is much better defined
for a $K$-band sample than in the optical. This is again because the
spectral slope in the optical is dominated by the star-formation rate;
thus spiral and elliptical galaxies have blue K-corrections that differ by 1
magnitude at $z=0.5$.  The observed morphological mix will therefore
change greatly with redshift, complicating the interpretation. In
contrast galaxy colours in the near-infrared are dominated by old stars
and are uniform across Hubble types (Aaronson 1978), thus yielding
a constant morphological mix.

A no-evolution prediction showed that we expected to see galaxies out
to $z=1$ at $K=17$ (Paper I), the redshift of interest for evolution,
so this motivated us to carry out a redshift survey of our
Paper I objects. Additionally the colour-redshift relation allows
us to test generic properties of spectral evolution models 
independently of the luminosity function.

The plan of this paper is as follows. Section 2 discusses the
spectroscopic observations and the data reduction procedures used to
obtain the redshifts and Section 3 describes how we use these to obtain
revised $K$ magnitudes in metric apertures for our galaxies.  In
Section 4 we discuss the details of the faint galaxy colours and how
they depend on redshift.  Section 5 details our luminosity function
analysis and in Section 6 we compare our results with other work on
galaxy evolution. Finally the results and conclusions are summarized in
Section 7.

Throughout, we scale results to the usual dimensionless Hubble parameter:
$h\equiv H_0/100\;\rm km\,s^{-1}Mpc^{-1}$. Unless otherwise stated,
we assume a cosmological model with $\Omega=1$ and zero cosmological constant.

\sec{SPECTROSCOPIC OBSERVATIONS}

Our redshift survey was carried out
in several observing runs on the Anglo-Australian Telescope and the
William Herschel Telescope in La Palma over the period 1990--1992.  We
used the  Autofib multi-fibre spectrograph (Parry \& Sharples,
1988) for the brighter objects ($R<19$), 
long and multi-slit spectroscopy
using the ISIS spectrograph for intermediate magnitudes ($18<R<20$)
and the LDSS1
(Wynn \& Worswick 1988) and LDSS2 (Allington-Smith \etal\  1994)
multi-slit spectrographs (both of which have a very similar design)
for the very faintest objects with ($R\gs 20$).

We followed standard procedures for debiasing, flatfielding and
stacking the  data and extracting spectra. Our final spectra were
determined to be limited by Poisson sky noise. For our faintest objects
at $R=21$--22, 9000s integrations were required to give $>90\%$
completeness in identifications. We determined the identifications
manually by carefully examing each individual spectrum, and also
used cross-correlation with galaxy templates following Tonry \& Davis (1979).
This confirmed our manual results, but proved no more powerful
at extracting redshifts from faint noisy spectra than the manual method.
In our initial spectroscopic runs we observed objects which we had
classified as stars from the image profiles in our broad-band CCD images;
it turned out our classification was very reliable (for the statistics
see Paper I) and so we subsequently observed only objects classified as
galaxies.
Our final redshift catalogue consists of 124 galaxies and is presented
in Table~A1 of the appendix.

As a result of an observing programme which evolved through a
succession of spectrographs of increasing sensitivity, our
dataset was acquired in a somewhat heterogeneous manner.
An ideal approach would have been to define in advance a
target sample which was randomly selected from the parent
catalogue to contain uniform numbers of objects per $K$
magnitude bin, and continue observing until redshifts for
all the target objects were obtained. In practice, the targets
we could observe were limited by scheduling of observing runs
and by weather. The initial runs lacked sufficient
sensitivity to yield redshifts for the reddest objects, and
so there were initially many cases of inconclusive spectra.
As the survey progressed and more sensitive spectrographs
became available, we were able to obtain successful spectra
representative of the previous class of failures, although
we could not in all cases observe the identical objects.
Our final spectroscopic runs were therefore used to
ensure that the redshift sample was as statistically
representative as possible.
The $(R-K,K)$ colour-magnitude diagram was inspected, and 
targets were chosen randomly to fill in under-sampled
parts of the $(R-K,K)$ plane. 
The success of this strategy may be judged from
Figure~1a, where we compare the $(R-K,K)$ distribution of the
objects with redshifts with that for all the galaxies.
We are very close to 
the desired uniform sampling of $R-K$ at given $K$, 
down to $K=17.25$; fainter objects were not considered.
The only areas where the eye suggests a low sampling
are at $(K,R-K)\simeq (17.0,4.5)$
and $(K,R-K)\simeq (15.5,2.0)$. The former case reflects
the difficulty of obtaining spectra for very faint objects, and
is quantified by the colour-dependent weights discussed in Section 5.1.
Conversely, any suggestion of bias against the blue objects is merely a
random fluctuation. Most spectra come from our October
fields, and redshifts were obtained for the bluest
objects in these fields; however, the March fields contain 
a few galaxies that are bluer than any in the October fields.

Since not all spectra yielded a redshift,
it is important to be sure that the omitted objects do not
bias the results. Our early runs on less sensitive spectrographs
had many inconclusive spectra, but this mainly allowed us to
establish empirically the integration time
needed for an object with a given $R$-band magnitude.
For each spectroscopic
run, we therefore defined a target integration time based
on the $R$-band magnitude. For an the different instruments
involved, this corresponded to a limiting magnitude in
10,000s integration of approximately $R=19.0$ (Autofib);
$R=21.0$ (LDSS1); $R=20.8$ (ISIS); $R\gs 22.5$ (LDSS2).
This results in 16 objects which satisfied our
integration-time criteria based on their $R$ magnitude, but
for which no redshift was obtained. The locations of these
objects on the colour-magnitude plane are shown in Figure~1b.
In almost all cases, the objects were within 0.5 mag. of the
effective optical limit for the instrument involved, and so
it is plausible that they are simply low-s/n versions
of the successful spectra, a selection of which are shown in
Figure~A1 in the appendix. Moreover it is clear
from Figure~1b that there is no significant bias in colour
or magnitude with respect to either the spectroscopically
identified galaxies or the larger sample of image-classified galaxies.
Given this and the small number
of such objects, we believe it is reasonable to assume that
the redshift distribution is not biased by their omission
from the sample. Conversely, as in any spectoscopic sample,
there are also a small number of identified
galaxies {\em fainter} than our nominal limits: these are also shown
in Figure~1b. There are only 8 of these
(numbers 77, 118, 132, 190, 316, 333, 362, 364); again they are within 0.5
mags of our completeness limits and they lie in typical
locations in the $(K,z)$ diagram; their inclusion makes no 
significant difference to our subsequent analyses.

In summary, our spectroscopic sample contains 124 of the 335
galaxies in the imaging survey of Paper I. Most of our redshifts
(119) are from the October fields, which contain a total of
201 galaxies, and we therefore have 59\% sampling
in this region. Furthermore, the selection of objects has
been adjusted so as to give a representative coverage of the 
$(R-K,K)$ plane (apart from a known reduced sampling at
faint $K$). Our sample should be statistically representative
of the infrared galaxy population.

Figure~2 shows the fundamental data in the form of the
redshift-magnitude plane. It is interesting to note
that substantial redshifts are attained at relatively
bright magnitudes. This observation already anticipates
one of our principal conclusions: that `merger' models
which postulate a faint characteristic luminosity at
high redshift are difficult to reconcile with our data.

\sec{APERTURE CORRECTIONS}

An important issue not fully explored in Paper I or
by Gardner \etal\ (1993) is the issue of aperture
corrections to the data. The majority of our data
were measured through a 4-arcsecond diameter
aperture, although some (the March fields) used
8 arcseconds. An aperture of about 6 arcsec was used
for most of the Hawaii work (see Gardner 1992).
In their recent paper on the $K$-band
luminosity function, Mobasher \etal\ 
(1993) corrected all their data to a standard isophotal aperture
based on the optical light profiles of their galaxies.
The metric aperture involved varied, but was typically
20 -- 30 $h^{-1}$ kpc, which would only correspond to
our 4 arcsec aperture at redshifts of $z\simeq 1$. Our
magnitudes are thus systematically fainter than those
defined by other workers; how much difference does this make?

With redshifts secured we are now able to remeasure magnitudes through
metric apertures; we will use a standard aperture of 20 $h^{-1}\rm
\,kpc$, which should be close to total for most galaxies and is a
typical diameter for local measurements. For large apertures ($\gs
10''$) our $K$ magnitudes start to become unreliable due to noise and
flat-field effects (due to the small size of the IRCAM detector). We therefore
follow the $K$ growth curves out to where they turn over, or become too
noisy ($\Delta K<0.2$); beyond this we adopt the growth curve
for the same galaxy from the corresponding $R$ band CCD image. This is
much better defined out to very large apertures.
Figure~3 shows the difference between the original 4 arcsecond $K$ magnitudes and
the 20 $h^{-1}$ kpc magnitudes as a function of redshift for galaxies for
which a 20 $h^{-1}$ kpc magnitude is directly measurable from the growth
curves. As expected the low-redshift galaxies are systematically too
faint, by up to 1 mag.

We might expect the growth in metric luminosity to be
well parameterised following the
approach of Gunn \& Oke (1975)
as:
$$
L(<r)\propto r^\alpha.
$$
For brightest cluster galaxies, $\alpha\simeq 0.7$
(Schneider \etal\ 19
83), but lower values
are more appropriate for field galaxies in general. For the sample of
Mobasher \etal\  (1993), typical effective values are $\alpha\simeq0.4$,
and this is consistent with the Hawaii data (Gardner 1992). 
We would expect such a power-law profile to be valid provided it
is not assumed to hold over too large a range of scales.
Our maximum redshift is 0.8, and all but 2 have
$z>0.06$; the range of proper diameters corresponding to
our 4 arcsecond apertures is thus 3.2 to 16.5 $h^{-1}\rm \,kpc$.
Even over this large range, exponential profiles
($r_{\rm scale}=3 h^{-1} $kpc)
and $r^{1/4}$ profiles ($r_{\rm eff}=4 h^{-1}$ kpc)
deviate from the power-law model at only at
the $\simeq 0.2$ magnitude level. We plot the 
$r^{0.4}$ prediction in Figure~3; it is an excellent
parameterisation of the data.

Our final metric magnitudes in $K$ and $R$ are given in Table~A1.
We carried out luminosity function analyses using both these values
and those predicted from the 4-arcsec data using the $r^{0.4}$
growth curve. The results were indistinguishable, as expected
from the good agreement shown in Figure~3.

\sec{FAINT GALAXY COLOURS}

We begin by looking at the optical--infrared colours of
our data. This will allow us to assess the
mix of Hubble types in this sample, as well as to
test synthetic galaxy spectra, on which we will rely to
K-correct the data in the luminosity function analysis.

For the colours we measure $K$ and $R$ magnitudes independently
in 20 $h^{-1}$ kpc apertures as in Section 3.
Figure~4 shows the 
$R-K$ colours of the galaxies plotted against
redshift. Objects dominated by emission features are plotted with a
separate symbol. We compare the colours with those of the galaxy
templates we used for the number count predictions in Paper I
(from Rocca-Volmerange \& Guiderdoni 1988; RVG), plotting
the red envelope for a high-redshift 1-Gyr burst of
star formation, which should provide an upper limit
to the locus of elliptical galaxies. We also show the
intermediate colour Sc type and the Im type,
which should define the blue limit for galaxies
on the Hubble sequence. For a $K$-selected survey
the K-corrections are very similar (see Section 5.2)
for all types and so the morphological mix should
not change with redshift. It is evident from
Figure~4 that the mix, as defined by colour, is
indeed approximately unchanging and so unlike optical surveys
we are not biased against high-redshift red galaxies.

We also consider the more recent GISSEL models of
Bruzual \& Charlot (1993).
These  use a more up-to-date library
of stellar templates and compute with a more accurate isochrone
synthesis technique. Particularly important for
our application is that the infrared portions of the
spectra are based on much more detailed data than the work of RVG.
Figure~4 also shows the GISSEL version of the 1-Gyr burst,
which makes a similar prediction for the red envelope, but
is systematically slightly bluer than the RVG model.
The latter seems in practice to be in better agreement
with the data.

We note the existence of one object (ID \#96) with very extreme colours
($z=0.225$, $R-K=5.7$).  This object has an emission line
spectrum and at this redshift the $\rm Pa\alpha$ line lies in the $K$
window; could this contribute to the $K$ flux?
Given standard case B assumptions for hydrogen line ratios
(Hummer \& Storey 1987) $\rm Pa\alpha/H\beta=0.332$, and we estimate
from the observed $H\beta$ flux that the line flux from
$\rm Pa\alpha$ would be equivalent to $K=27$. 
There would therefore need to be large amounts of extinction
in order for line emission to be significant, but this is
not indicated in the optical spectrum. 
This object merits further study.

Otherwise, the reddest galaxies are approximately consistent
with the model red envelope to within observational
error, with the possible exceptions of \#109, \#224 \& \#346.
This was not the case with our first version of this
diagram for 4-arcsecond aperture colours, 
which contained many galaxies with $z<0.3$ much
redder than the envelope, particularly at low redshift.
This was initially puzzling, but it was eventually realized that this
was an effect of colour gradients:  at low redshifts, the 4-arcsecond
apertures sample the galaxy nuclei only -- and these are very red in
some cases.  The galaxies with particularly red nuclei, together with
their redshifts and 4-arcsecond colours are: \#224, \#563, \#406,
\#109, \#392 and \#334. $z=(0.063, 0.080, 0.121, 0.148, 0.153, 0.192)$,
$R-K=(3.8, 3.7, 3.9, 4.0, 3.9, 4.0)$.  Significant optical-infrared
colour gradients in ellipticals were previously noted by Peletier
\etal\  (1989). They find up to 0.6 mag of reddening in $V-K$ for a
factor 10 in radius, and our results seem to be consistent with these
more extreme values.

What is the cause of the red nuclei? Although galactic bulges are
redder than disks, we still would not expect them to be redder than a 
1-Gyr burst if the colours were due to the stellar populations.  It is
possible with the GISSEL code to choose various Initial Mass Functions,
and it is interesting to ask if the results are robust, particularly
for the red envelope.  We find that stars with initial masses
$>2.5\Msun$ make negligible contributions at late ages ($>15$ Gyr).
Increasing the proportion of low mass stars by adopting a flatter IMF
also makes little difference -- doubling the fraction of stars below
$0.3\,\Msun$ only makes a difference to the curve at the 0.05
magnitudes level. We conclude that the choice of IMF has little
influence on the red envelope of $R-K$ colours, as it is determined by
the spectral energy distributions of $\sim$ solar mass stars. Peletier
\etal\  attribute the reddening to metallicity gradients. This is
certainly a plausible explanation for our results, although large
amounts of dust in some nuclei remains a possibility
(a screen with $A_V=1.2$--$1.3$ mag. would be required to produce the
observed $R-K$ excess).  Low-level AGN are a third potential
explanation, especially if they are heavily reddened.
These extremely red galaxies merit detailed further 
examination to investigate these possibilities.

\sec{LUMINOSITY FUNCTION ANALYSIS}

We now proceed to derive the $K$-band galaxy
luminosity function from our data. We shall be
particularly interested in the comparison between
our results and those of Mobasher \etal\  (1993)
and the Hawaii survey (Cowie \& Songaila 1993;
Cowie \etal\  1995). The former was based on $K$-band
observations of 95 $B$-selected galaxies, with
completeness claimed to $K=12.5$. The Hawaii sample
consists of 262 redshifts, complete to $K\simeq 19$ -- 20.
In fact, our results turn out not to agree very
well with either of these pieces of work, for 
what we believe are the reasons described below
in Section 5.4.

\ssec{Counts and incompleteness corrections}

The longer integration times required mean that 
we were able to obtain fewer spectra for the fainter galaxies.
Nevertheless, the redshift distribution at given
magnitude should still be faithfully reproduced
by our data, provided the range of colours at
given $K$ is properly sampled. We tested this by
dividing the $(R-K)$ -- $K$ colour--magnitude plane
into cells (using increments of 0.3 in $K$
and 0.5 in $R-K$) and comparing the populations of these
cells in our total and spectroscopic samples.
This allows a colour-dependent weight to be
deduced at given $K$ magnitude. These weights were
usually close to unity; setting them to exactly unity
had no significant effect on the results below ---
the maximum change in $M^*_K$ is only 0.02 mags and
the effect on the resultant space densities is at
the $\ls 10\%$ level. As the latter spans a factor
of 100 in value it is not suprising that $M_K^*$
is robust to such changes.

We now need to know the effective $K$-dependent
incompleteness, and this may be deduced 
by comparing the number of galaxies
in our spectroscopic sample as a function of magnitude
with that expected from the overall number counts.
A convenient analytical fit for these is
$$
{dN\over dK} \, / \, {\rm deg}^{-2} =
{ 10^{0.75(K-12.1)} \over
\left[ 1 + 10^{0.35(K-17.2)}\right]^{1.5} },
$$
which is a statistically acceptable best fit to the
data from Paper I plus the Hawaii counts from
Gardner, Cowie \& Wainscoat (1993), and 
the data of Jenkins \& Reid (1991), as shown in Figure~5
(this plot also includes the recent faint count data from
McLeod et al. 1995 and Djorgovski et al. 1995).
We have preferred to force the
slope at bright magnitudes to the Euclidean value
expected from the aperture correction:
$N\propto 10^{\beta K}$, where $\beta=1.2/(2-\alpha)$ (where $L\propto r^{\alpha}$
is the analytic aperture correction derived in Section 3); 
any other slope would indicate
strong local evolution if taken literally.

The corrections needed
to achieve uniform 4 arcsecond magnitudes in Figure 5 are as follows.
The Hawaii data (Gardner 1992)
were measured in 6.3 arcsec apertures for the HMDS and
HMWS surveys, but published with offsets of respectively
$-$0.1 and $-$0.2 mags as a notional correction to total.
We have removed these offsets and added a further
0.2 mag to correct to 4-arcsec measurements.
The deeper HDS was measured in 3.5 arcsec and a correction
to 6 arcsec made on a field-to field basis. Since the individual
corrections are not available, we have treated the published
data as exact 6-arcsec measurements and added a correction of
0.18 mag.
The deep data of Djorgovski et al. (1995) are in 5.4 arcsec
apertures, and so need a correction of 0.13 mag.
The deep data of McLeod et al. (1995) are focas magnitudes
for which an aperture is not quoted; we have left them uncorrected.
Lastly, for the 20 arcsec apertures used by Jenkins \& Reid (1991),
the correction to 4 arcsec predicted by our $r^{0.4}$
growth curve is 0.70 mag. 
This explains why their counts were clearly seen in Paper I
to lie above those obtained
by other workers. However, we do not expect a power law to
apply over this range of radius; at faint magnitudes, apertures
above 8 arcsec will be total. We have therefore applied the offset
of 0.7 at $K=15$, declining to 0.3 at $K=19$. This should
be correct to within about 0.1 mag.

Dividing the $K$ counts in our spectroscopic sample
by the average counts yields the effective completeness,
shown in Figure 6. This stays close to unity up till
about $K=16$ and then falls to almost 0.1 in the
$K=17$ -- 17.25 bin, which is the faintest bin that
contains spectroscopic data. The completeness is
significantly greater than 1 for $K\simeq 14.5$, and
this reflects the use of some galaxies as positional
references in constructing the sample (see Paper I).
We shall use this incompleteness curve in the luminosity
function analysis, assuming unit completeness for
$K<14$. Using a smoothed form of this figure makes
no difference to the results.

We can now deduce the properties of a true flux-limited
sample to $K=17.25$ by appropriately weighting the
fainter galaxies. Figure 7 shows both the raw and
weighted redshift histograms for our sample. The
observed median redshift of 0.24 increases to 0.35 after
weighting.

\ssec{K-corrections}

In order to obtain absolute magnitudes,
we require a knowledge of the luminosity distance $D_{\ss L}$,
the $K$-band K-correction $K(z)$, and the aperture correction $A(z)$:
$$
M(z)=m-5\log_{10}[D_{\ss L}/10\;{\rm pc}]-K(z)+A(z).
$$
For simplicity, we shall throughout quote absolute magnitudes 
assuming $h=1$ for the Hubble parameter.
The aperture correction converts
the observed aperture magnitudes to some proper diameter.
As stated above, we shall choose this to be 
$$
D_0= 20h^{-1}\;\rm kpc,
$$
so that the aperture correction is given in terms of
angular-diameter distance $D_{\ss A}(z)$ as
$$
A(z)=\log_{10}[\theta D_{\ss A}(z)/ 20h^{-1}\; \rm kpc],
$$
where $\theta$ is our standard angular diameter of 4 arcsec.

One of the advantages of the infrared waveband is that the 
K-corrections are very similar for all classes of galaxy, 
reflecting the dominance of giants in this waveband. The widely 
different amounts of star formation in different Hubble types 
only affects the spectra at wavelengths somewhat shorter than 
1$\mu$m. This is illustrated in Figure 8, which shows 
theoretical K-corrections taken from the evolutionary synthesis 
models of Bruzual \& Charlot (1993; BC). Rather than a range of
models designed to fit the Hubble sequence, we show an
instantaneous burst of star formation observed at ages
from 1 to 10 Gyr. There is satisfyingly little model dependence
of the K-corrections. 
The following is a good fit to the 5-Gyr data
for $z\ls 1.5$:
$$
K(z)={-2.58z+6.67z^2-5.73z^3-0.42z^4 \over
1-2.36z+3.82z^2-3.53z^3+3.35z^4},
$$
and we use this as our standard K-correction. 

For comparison, we also show the K-correction for
the `UV-hot' elliptical model of
Rocca-Volmerange \& Guiderdoni  (1988; RVG). 
A good fit to their data for $z\ls 1.5$ is
$$
K(z)=-(1+[5z]^{-3/2})^{-2/3}.
$$
The BC K-corrections show more structure than those
from RVG, reflecting the less sophisticated treatment of 
infrared wavelengths by RVG.
There is also a systematic
difference between the BC and RVG models, of about 0.2 mag
at $z=0.5$, in the sense of RVG being bluer (although their
models are redder in $R-K$).
Our absolute magnitudes would thus be fainter at high
redshift by this amount if we adopted the GRV K-correction.
However, we are confident that the BC relation is more
nearly correct, since it accounts well for the $JHK$
colours of local galaxies (Aaronson 1978; Mobasher \etal\ 1993).

This completes the ingredients needed to deduce
absolute magnitudes. We therefore show in Figure 9 the raw
data for the luminosity function analysis: the
area of the luminosity-redshift plane sampled
by our survey. The following Sections analyze this
distribution in order to deduce space densities.

\ssec{Luminosity function estimates}

The simplest estimator of the luminosity function is to bin up 
the data in redshift slices as a function of absolute 
magnitude. The estimator for the density in a given bin is then 
the traditional
$$
\hat\phi=\sum_i\hat\phi_i=\sum_i {w_i\over V(z_{\rm max})-V(z_{\rm min})}
$$
(Felten 1976), where $z_{\rm max}$ is the smaller of the 
maximum redshift within which a given object could have been 
seen, and the upper limit of the redshift band under 
consideration; $z_{\rm min}$ is the lower limit of the band. In 
this case, the weight to use is the full product of corrections 
for colour-dependent incompleteness and reduced sampling at 
faint $K$. The result is shown in Figure 10a, for redshift bins 0 
-- 0.2, 0.2 -- 0.4, 0.4 -- 0.8. 

An alternative way of presenting the same data has been favoured
by the Hawaii group, which is to use the cumulative luminosity
density. The obvious estimator for this is
$$
\hat\rho(>L)=\sum_{L_i>L} L_i\, \hat\phi_i,
$$
and the results are shown in Figure 10b, where luminosities
have been converted to solar units on the assumption that
the solar luminosity corresponds to $M_K(\odot)=3.4$.

In both cases the message is the same, although the
cumulative estimator appears (perhaps misleadingly) less noisy.
While the two low-$z$ slices are very similar, it
is clear that the characteristic luminosity is higher in the
$0.4<z<0.8$ slice, by at least 0.5 mag. It also seems as though
the overall luminosity density is very nearly constant.

We now quantify these visual impressions by model fitting.
It is convenient to describe the galaxy luminosity function
via a Schechter function fit at each redshift
$$
d\phi = 0.921\,\phi^* (L/L^*)^{\alpha+1} \exp[-L/L^*]\; dM.
$$
The optimal way of fitting such models to moderate discrete datasets
such as ours is to use maximum likelihood. 
In the absence of clustering, one would define likelihood by
$$
{\cal L}=\prod_i {d^2p\over dM\,dz}(M_i,z_i),
$$ 
and  extra constraints such as operating over a redshift
band can be applied by restricting the product to the
relevant objects and normalizing the model probability
distribution to the required region of ($M,z$) space.

The presence of clustering renders the vertical normalization of the
luminosity function uncertain. It may also affect the shape of the
function, but such luminosity segregation has never been demonstrated
convincingly, and we shall assume here that it is smaller than our
statistical errors. It is unclear how good this
assumption is in the infrared: the known phenomenon of
morphology segregation plus the tendency for ellipticals to
be redder than spirals should produce some brightwards
shift in characteristic luminosity -- so that a positive density
perturbation boosts the number density of bright galaxies in two ways.
The following method at least avoids the direct density boost,
and so should be closer to the average luminosity
function. In the end, there is no substitute for an
area which is large enough to be representative.

The above method can now
be applied directly for an infinitesimal redshift band, since
only the probability distribution for $M$ at given $z$ is
involved and amplitude scalings normalize away:
$$
{\cal L}=\prod_i {dp\over dM}(M_i \mid z_i).
$$
This expression can be immediately generalized to a finite
redshift range by continuing to use the conditional
probability of $M$ at given $z$ -- but this must now
be normalized individually for each $z_i$ of interest.

A last problem is how to deal with incompleteness.
We have deduced a set of weights $w_i$ which account for
the sampling incompleteness associated with each object,
so it is tempting to modify the likelihood to
$$
{\cal L}=\prod_i \left[ {dp\over dM}(M_i \mid z_i) \right]^{w_i}
$$
(e.g. Zucca \etal\  1994).
However, although this would eliminate gross biases in the
answer, it is clearly not satisfactory statistically. This expression
corresponds to counting a few faint objects many times, so
that the error bars will be characteristic of a 
larger sample than the real one -- i.e. they will be 
spuriously small. The correct approach is to incorporate
the incompleteness into the model:
$$
{\cal L} = \prod_i {  \phi(M_i,z_i) \, C(M_i,z_i)  \over
               \int_{-\infty}^{\infty} \phi(M,z_i) \, C(M,z_i) \, dM }
$$
where the completeness factor $C$ accounts for sampling factors
and magnitude limits, 
and is incorporated into the normalisation.
It is easy
to deal with our $K$-dependent sampling in this way.
Any colour dependence is harder to deal with, however, since
this does not have a direct relation with $M$ \& $z$. We
therefore used the ${\cal L}\propto p^w$ prescription for
the colour weights only. Since these are unity on average
and have a small deviation from unity, the fact that this
is not formally the correct procedure will not be a problem
in practice. In fact, setting all colour weights to unity
has no significant effect on our results.

This method gives a value for the characteristic luminosity in
a redshift band, $L^*(z)$; the normalization $\phi^*(z)$ can then
be determined from the overall numbers of objects (although
it is still subject to clustering fluctuations).
The errors quoted below assume that luminosity density
can be measured exactly, so that the fractional error
on $\phi^*$ is the same as that on $L^*$.
The results of the analysis are given in Table~1, assuming
$\Omega=1$ and a Schechter-function slope of $\alpha=-1$
(letting this float yielded a best-fitting value of
$\alpha=-1.04\pm0.31$).

\vbox{\strut\hfill\break
\noindent
{\bf Table 1.} Luminosity function fits
\bigskip
\centerline{\vbox{
\tabskip 0.5truecm
\halign{\hfil#\hfil &\hfil#\hfil &\hfil#\hfil \cr
z & $M_K^*(z)$ & $\phi^*(z)/h^3\;\rm Mpc^{-3}$ \cr
\cr
0.0 -- 0.2 & $-22.72 \pm 0.23$ & $0.029\pm 0.007$ \cr
0.2 -- 0.4 & $-22.85 \pm 0.17$ & $0.020\pm 0.003$ \cr
0.4 -- 0.6 & $-23.23 \pm 0.23$ & $0.013\pm 0.003$ \cr
0.6 -- 0.8 & $-23.68 \pm 0.30$ & $0.009\pm 0.002$ \cr
\cr
0.0 -- 0.4 & $-22.75 \pm 0.13$ & $0.026\pm 0.003$ \cr
0.4 -- 0.8 & $-23.41 \pm 0.24$ & $0.011\pm 0.003$ \cr
\cr
0.0 -- 0.8 & $-23.01 \pm 0.11$ & $0.019\pm 0.002$ \cr
}
}
}
\bigskip
}

These numbers paint an interesting picture, and
confirm earlier visual impressions. There indeed appears
to be some evidence for luminosity evolution in the sense
that $M_K^*$ was brighter in the past. The no-evolution 
hypothesis is ruled out at about the 4 per cent significance
level, considering the variation of $M^*_K$ alone.
On the other hand, there is no evidence for evolution
for $z<0.6$. Furthermore, there is evidence that the
overall normalization of the luminosity function is a
declining function of redshift.

A simple model that accounts for what is seen is
therefore to take the low-$z$ parameters for the luminosity function
$$
M_K^*(0)=-22.8;\quad \phi^*(0)=0.026h^3\;{\rm Mpc}^{-3},
$$
and scale them approximately as $L^*\propto (1+z)$ and $\phi^*\propto (1+z)^{-1}$.
This looks very like the merging models advocated by
Broadhurst \etal\  (1992), with an approximately
conserved luminosity density -- except that the evolution is
in the opposite sense.

These results are derived on the assumption of
an Einstein-de Sitter model. Since we have made aperture
corrections that involve fixed metric diameters, the
whole analysis should in principle be re-done 
from the start for
any different model. However, since the aperture corrections
are only important at low redshift, it will suffice
to ask how the luminosity distance and volume element
change for different models. To illustrate the
model sensitivity, we focus on $z=0.7$, which is the
centre of our highest-$z$ bin. We consider two
popular alternative models: (A) an open universe with
$\Omega=0.2$; (B) a $k=0$ model with $\Omega=0.2$ in
matter and $\Omega=0.8$ in vacuum energy. 
The required distances and volumes,
divided by those for the Einstein-de Sitter
model, are $D_{\ss L}(0.7)= 1.14$ (A) and 1.32 (B);
$dV(0.7)= 1.60$ (A) and 2.89 (B).
Adoption of these models would thus exacerbate
the trends we have identified: $M^*$ at high redshift
would be 0.3 -- 0.6 mag. brighter, and $\phi^*$
would be a factor 1.6 -- 2.9 lower. Note that
the total luminosity density would decline
only slightly: this is a relatively robust quantity.

Lastly, we consider the question of dependence of the
luminosity function on colour. Using the colour-redshift
plot of Figure~4, it is possible to divide the sample
at approximate Hubble-type boundaries
by simple vertical shifting of any of the model lines.
We have partitioned the sample into three
equal parts in this way, and galaxies of different
colour are indicated by different symbols in Figure~9.
Restricting attention to $z<0.5$, where the overall
sample has no evidence for evolution, we find the
following $M^*$ values (assuming $\alpha=-1$):
$-22.85\pm0.18$ (reddest: approximately E/Sa);
$-22.92\pm0.18$ (intermediate: approximately Sb);
$-21.32\pm0.28$ (bluest: approximately Sc/Im).
There is thus a strong trend for the bluest
galaxies to be less luminous.
However, in agreement with
Mobasher \etal\ (1993), we find little difference between
$M^*$ for the two reddest categories: 
ellipticals and early-type spirals.

\ssec{Comparison to other results}

These numbers are very different from the results of
Mobasher \etal\  (1993), who (for $h=1$) obtained
$M^*=-23.6\pm 0.3$ and $\phi^*=0.0046\pm0.0011$.
How can it be that we have obtained a characteristic luminosity
a magnitude fainter and a normalization over 5 times higher?
It sounds like there is some error, but the numbers are
not as different as they seem. Mobasher \etal\  used
isophotal magnitudes, rather than a fixed metric aperture.
If we consider their objects with $z\simeq 0.1$ (the highest
redshift objects, which are the most luminous and which
thus dominate the determination of $M^*$), their median
aperture is approximately 40 $h^{-1}$ kpc diameter, which
immediately makes their magnitudes 0.3 mag. brighter than
ours, if we adhere to the power-law aperture correction. 
Also, the K-corrections used are different: they
adopt $K_K(z)=-0.7z+3.9z^2$, a much weaker dependence than
our $K_K(z)\simeq-2.58z$. At $z=0.1$, the difference in K-correction
is 0.22, so that this plus the aperture difference accounts
for 0.52 mag. of the 0.8 mag. difference in $M^*$; the remaining
difference is not statistically significant. As for the
difference in $\phi^*$, this may well be partly due to
density fluctuations, but it is also possible that
the Mobasher \etal\  sample is systematically incomplete:
since their data were based on infrared measurements of
blue-selected galaxies, the population of very red
nearby galaxies will not be sampled adequately. If the
Mobasher \etal\  sample is incomplete for faint $K$, this
would produce a spuriously low normalization and a spuriously
bright $M^*$. Mobasher \etal\  used a $V/V_{\rm max}$ test to
assess completeness, but they have a very rich cluster at
$z\simeq 0.04$, and so this test for completeness is
invalid, since it relies on spatial homogeneity.
It is also possible that any tendency towards brighter $L^*$
in clusters might bias their result, although they failed
to detect any systematic difference in the luminosity function
for ellipticals.
An alternative viewpoint is to worry that our $L^*$ may
be too faint because our `blank-field' survey regions were
necessarily chosen free of extremely bright objects.
There are two arguments against this being important:
(i) we have the same fainter $L^*$ in the $z=0.2$ -- 0.4 bin,
where the brightest objects are fainter than the positional references;
(ii) luminosity function fitting to the binned data ignoring
the existence of empty bright bins gives consistent $L^*$ values.

Similarly, our results diverge quite markedly from those
of the Hawaii group (Cowie \etal\ 1995).
They obtain a total $M_K^*\simeq-23.4$, which is claimed
not to evolve, and a normalization which changes approximately
as $(1+z)^2$. The local value of $\phi^*$ for their data
is not quoted, but is approximately $0.006h^3$ Mpc$^{-3}$.
This difference between our results and those of the Hawaii group
is more disturbing, since they have a much larger area than
Mobasher \etal, and almost twice as many redshifts as we do.
On the other hand, their sampling rate declines rapidly
with $K$: in the region containing most of our data ($15<K<17$),
we have 103 redshifts, whereas they have 73. The fact
that we see a clear increase in $L^*$ at high redshifts, whereas
they do not, is barely consistent with limited statistics.

The difference is clearly in the raw data and {\em not\/} in
the analysis: we have analysed their dataset using the
method outlined in Section~5.3 and obtained identical
results to Cowie \etal\
Much turns on the status of the rare bright galaxies at
high redshift. We have 9 objects with $z>0.6$, $K<17$,
whereas the Hawaii group have 5. Either they are missing a
few objects, or we have an upward fluctuation. However, note
that our method of analysis should be robust with respect to
density fluctuations. One might worry about having a single
rich cluster in the survey with an atypically bright $L^*$,
and we do indeed have some kind of enhancement in number at
$z\simeq 0.65$. However, the galaxies here come from two
widely separated fields, and there is only one pair within
a proper separation of 1 $h^{-1}$ Mpc. It therefore seems
implausible that our bright $L^*$ at $z>0.6$ can be biased
by density fluctuations.
However, a larger sample is clearly required
for a definitive answer.

Most odd of all is the large discrepancy in luminosity density
at low redshifts between our results and those of the Hawaii group. We
note that in their lowest redshift bin, $0<z<0.2$, Cowie \etal\ have a
rather low normalisation for their luminosity function, equivalent
to $\phi^*\simeq 0.006 h^3$ Mpc$^{-3}$. Most workers estimate the local
optical $\phi^*$ to lie in the range $0.01-0.03 h^3$ Mpc$^{-3}$ (e.g. Loveday
\etal's (1992) luminosity function analysis);
these values are approximately 2 -- 5 times larger.
We therefore suspect that the lowest-redshift bin of the
Hawaii data may be incomplete. If this is ignored, the
difference between our higher-redshift results is not so great, 
as discussed above.

\sec{IMPLICATIONS FOR FAINT COUNTS}

The obvious question now is how the models which fit the
data at $K\ls 17$ and $z\ls 1$ will fare when extrapolated
to fainter $K$ magnitudes and higher redshifts. Figure~11
shows the number-count data compared to selected models.
The interesting thing here is how well the no-evolution
model fits the data, contrary to previous claims.
As discussed above, these were made on the basis of
analyses which ignored substantial aperture corrections.
The counts fall below the Euclidean prediction at about
the point where $M^*$ reaches $z=1$; since we now have a
somewhat fainter $M^*$, the turnover moves to fainter
magnitudes and matches well the observed decline of the
counts. Our increased $\phi^*$ value means that the overall
level of the counts are correctly predicted, as well as the shape.
Since our redshift data only rule out the
no-evolution model at a moderate level of significance,
it seems that this is something that should still be
taken seriously. Moving to pure $L^*\propto 1+z$
luminosity evolution with the same local normalization 
now significantly exceeds the faint
counts, whereas it was previously claimed to give a good
match. However, conserving luminosity density by
scaling $\phi^*\propto (1+z)^{-1}$ at high redshifts
restores the good fit at most magnitudes. The predicted
numbers are too small at $K>21$, but this should not be
very surprising, since the typical luminosities of galaxies
at that level are fainter than we have been able to probe
in our luminosity function determination.
One simple possibility is that the luminosity function has
an extra dwarf component which makes it steeper at the
faint end; several authors have argued for such a component
both in cluster luminosity functions (e.g. Driver et al. 1994)
and in the field (e.g. Gronwall \& Koo 1995).
Driver et al. obtain the following parameters for the
dwarf luminosity function:
$$
\eqalign{
M^*_{\rm dwarf} &= M^*_{\rm normal} + 3.5 \cr
\phi^*_{\rm dwarf} &= 2\phi^*_{\rm normal}\cr
\alpha_{\rm dwarf} &=-1.8.\cr}
$$
Adding such a (non-evolving) low-luminosity contribution to our
antimerging normal luminosity function fits the faint
counts with no adjustment of parameters.
We therefore suggest that this combination be regarded as a
`standard model' for the infrared luminosity function.

Can these possibilities be constrained by fainter redshift
data? Cowie \etal\ (1994) have described redshift surveys
as faint as $K(6'')=20$, and shown that the median redshift
continues to follow their no-evolution prediction
down to $K(6'')=18$ -- 19 ($z_{\rm med}=0.65\pm0.15$), but
to diverge at
$K(6'')=19$ -- 20 ($z_{\rm med}=0.5\pm0.2$).
Our predictions for these median redshifts are
0.70 \& 0.97 respectively for the no-evolution model,
0.87 \& 1.17 respectively for the luminosity/density evolution model, and
0.84 \& 1.02 respectively for the luminosity/density evolution model
with extra dwarfs. The departure from all
these models in Cowie \etal's
faint $K=19$--20 bin is rather severe. 
In the context of the model that includes dwarfs, 
we note that Cowie et al.'s data do contain the suggestion of
a low-luminosity clump around $K=19.5$ -- 20, $z\simeq 0.2$,
which is where the dwarf population would first manifest itself.
However, more faint redshift data are really required
for a definite statement:
the true median redshift in Cowie \etal's
faintest bin may well be higher than the figure they
estimate from a set of 22 galaxies, of which 9 have
redshifts estimated from colours.

\sec{CONCLUSIONS}

We have presented an unbiased infrared-selected redshift
survey of 124 galaxies to $K=17.25$ and deduced the
evolution of the $K$-band luminosity function.
Our principal conclusions are

\item{(i)} That the local normalization of the luminosity function
is somewhat higher than has been found in previous work:
$\phi^*=0.026\pm0.003\, h^3\, {\rm Mpc^{-3}}$. 

\item{(ii)}
Combined with a slightly fainter $L^*$ than previous work,
we find that an $\Omega=1$ model with no evolution fits
the number counts rather well to $K=21$.

\item{(iii)}
However, our data indicate that $L^*$ is brighter at high redshift,
by at least 0.5 mag. at $z=0.7$.

\item{(iv)}
Positive luminosity evolution in this sense then fits the
counts only if (a) there is corresponding negative density
evolution and (b) there is an additional dwarf component
at very low luminosities.

How do we relate these results to optical studies of
the galaxy population? We are unable to say much about the
`faint blue galaxies' that dominate the faint 
optical counts: given their colours, we would expect them to have
$M_K\sim -21.5$, and at $K<17.25$ we would not be able to see them
beyond $z>0.15$. Rather they would not manifest themselves until
$K=19$--20, and so this suggests an obvious connection between the
faint excess in the blue counts and the possible dwarf population
discussed above. Such an idea is given further support by the
results of Glazebrook et al. (1995b), based on HST imaging
of random galaxy fields. They find counts of morphologically
normal ellipticals and spirals to be much as expected from
no-evolution models, whereas the faint excess arises from
steep number counts in the irregular population.

Nevertheless, because we favour a model in which the
normalization is high, this does have implications for
evolution in the optical. Various workers (e.g. Metcalfe \etal\ 1991; 1994)
have argued for a high normalization, based on the
good fit of a no-evolution model around $b_{\ss J}=19$.
This would then imply that the bright counts
are incomplete, by a factor of 2 at $b_{\ss J}=17$, favouring
a prosaic explanation such as that of McGaugh (1994).
In this sense, the implication of our result is that 
the faint blue galaxies may be less dominant than often
supposed, and less in need of radical explanations.

Turning to the infrared luminosity function, what are
the physical implications of the evolution we have detected?
Our results stand in
direct contradiction to the simple scaling merger model of BEG, and
imply that at least the most luminous galaxies may have had a
relatively uneventful history. The amount of luminosity evolution 
we see is consistent with what is inevitably expected from
passive aging of stellar populations -- a point stressed
by Cowie \etal 1995. The evolution of the bright end
is also similar to that seen in 3CR
radio ellipticals by Lilly \& Longair (1984) --
although see Dunlop \& Peacock (1993) for evidence that
the infrared light in 3CR galaxies probably has an AGN-related component.

The bright end of the luminosity function is therefore
consistent with a picture in which
massive spheroids were in place at $z\gs 1$ and
have subsequently evolved passively (see e.g. Bower \etal\ 1992
for other pieces of evidence in favour of this conclusion).
However, we do not see a luminosity density which declines
with time, as expected from passive aging alone.
A possible interpretation of this fact is that additional
star formation at intermediate redshifts enhances the
luminosity function of lower-luminosity galaxies. In
a picture where spheroids are old and passively evolving,
it would be natural to associate this feature with
the epoch of disk formation (e.g. Gunn 1982).
The failure of this process to affect the bright end of the
luminosity function would then be related naturally to
morphological segregation and the preference of the
most luminous ellipticals for dense environments.

Clearly, this is only a preliminary and 
qualitative picture, which requires
further testing and refinement. Of particular
interest will be the evolution at $z>1$, which is rather poorly
constrained by existing data. This remains an outstanding task
for the new generation of large telescopes.

\AutoNumberfalse

\sec{ACKNOWLEDGEMENTS}

\tx
We acknowledge the generous allocations of telescope time on the U.K Infrared
Telescope, operated by the Royal Observatory Edinburgh, the Isaac Newton
Telescope, operated by the Royal Greenwich Observatory in the Spanish
Observatorio del Roque de Los Muchachos of the Instituto de Astrofisica de
Canarias and the Anglo-Australan Telescope.  We also thank the staff and telescope operators of these telescopes
for their enthusiasm and competent support.
The photographic photometry was performed on plates supplied by the
U.K. Schmidt Telescope Unit using the COSMOS measuring machine at the Royal
Observatory, Edinburgh.  The computing and data reduction was carried out on
STARLINK which was funded by the SERC. Special thanks go to James Dunlop
for allowing us to use his software to 
compute the K-corrections from galaxy templates, and to
Len Cowie for an electronic version of the Hawaii data.
KGB acknowledges the support of a SERC
research studentship.

\sec{REFERENCES}



\def\name#1{#1}     
\def\vol#1{#1}      
\def\MNRAS{MNRAS}
\def\ApJ{ApJ}
\def\AstrAst{A\&A}
\def\AstrAstSup{A\&AS}
\def\AstrJ{AJ}

\ref Aaronson M., 1978, \name{\ApJ}, \vol{221}, L103

\ref Allington-Smith J. R., Breare J. M., Ellis R. S., Gellatly D. W., Glazebrook
K., Jorden P. R., MacLean J. F., Oates A. P., Shaw G. D., Tanvir N. R.
Taylor, K., Taylor P. B., Webster J., Worswick S. P., 1994,
\name{PASP}, in press

\ref Babul A., Rees M. J., 1992, \name{\MNRAS}, \vol{255}, 346

\ref Bower R.G., Lucey J.R., Ellis R.S., 1992, \name{\MNRAS}, \vol{254}, 601

\ref Broadhurst T. J., Ellis R. S., Shanks T., 1988, \name{\MNRAS}, \vol{235}, 827

\ref Broadhurst T. J., Ellis R. S., Glazebrook K., 1992, \name{Nat}, \vol{355}, 55 (BEG)

\ref Bruzual A. G., Charlot S., 1993, \name{\ApJ}, \vol{405}, 538

\ref Carlberg R. G., 1992, \name{\ApJ}, \vol{399}, L31

\ref Colless M. M., Ellis R. S., Taylor K., Shaw G., 1991, \name{\MNRAS}, \vol{253}, 686

\ref Cowie L. L., Songaila A., Hu E. M., 1991, \name{Nat}, \vol{354}, 460

\ref Cowie L. L., Songaila A., 1993, proc 8th IAP meeting, eds B. Rocca-Volmerange et al., Editions Frontieres, Paris, p147. 

\ref Cowie L. L., Gardner J. P., Hu E. M., Songaila A., Hodapp K. W., Wainscoat R. J., 1994,  \name{\ApJ}, \vol{434}, 114

\ref Cowie L. L., Songaila A., Hu E. M., 1995,  \name{\ApJ}, in press

\ref Djorgovski, S. et al. 1995, \name{\ApJ}, \vol{438}, L13.

\ref Driver, S.P., Phillipps, S., Davies, J.I., Morgan, I., Disney, M.J., 1994, \name{\MNRAS}, \vol{268}, 404.

\ref Dunlop J.S., Peacock J.A., 1993, \name{\MNRAS}, \vol{263}, 936.

\ref Ellis R. S., 1990, in Kron R. G., ed, \name{Evolution of the Universe of
Galaxies}, A.S.P.\ Conference Series, Vol.10, p248

\ref Felten J. E., 1976, \name{\ApJ}, \vol{207}, 700

\ref Gardner J. P., 1992, PhD thesis, University of Hawaii,

\ref Gardner J. P., Cowie L. L., Wainscoat R. J., 1993, \name{\ApJ},
\vol{415}, L9

\ref Glazebrook K., Peacock J. A., Collins C. A., Miller L., 1994,
\name{\MNRAS}, \vol{266}, 65

\ref Glazebrook K., Ellis R. S., Colless M. M, Broadhurst T. J., Allington-Smith
J. R., Tanvir N. R., Taylor K., 1995a, \name{\MNRAS}, in press

\ref Glazebrook K., et al., 1995b, \name{Nat}, submitted

\ref Gronwall C., Koo D.C., 1995, \name{\ApJ}, \vol{440}, L1

\ref Gunn J. E., Oke J. B., 1975, \name{\ApJ}, \vol{195}, 255

\ref Gunn J. E., 1982, proc. Vatican conference on Astrophysical Cosmology, eds H.A. Br\"uck et al., 
Pontificae Academicae Scripta Varia, p233.

\ref Hummer D. G., Storey P. J., 1987, \name{\MNRAS}, \vol{224}, 801

\ref Jenkins C. R., Reid I. N., 1991, \name{\AstrJ}, \vol{101}, 1595

\ref Koo D. C., Gronwall C., Bruzual G. A., 1993, \name{\ApJ}, \vol{415}, L21

\ref Lilly S. J., Longair M. S., 1984, \name{\MNRAS}, \vol{211}, 833

\ref Loveday J., Peterson B. A., Efstathiou G., Maddox, S. J., 1992,
\name{\ApJ}, \vol{390}, 338

\ref McLeod B.A., Bernstein G.M., Rieke M.J., Tollestrup E.V., Fazio G.G., 1995,
\name{ApJ Suppl.}, \vol{96}, 117

\ref McGaugh S.S., 1994, \name{Nat}, \vol{367}, 538

\ref Metcalfe N., Shanks T., Fong R., Jones L. R., 1991, \name{\MNRAS}, \vol{249}, 498

\ref Metcalfe N., Shanks T., Fong R., Roche N., 1994, \name{\MNRAS}, submitted

\ref Mobasher B., Ellis R. S., Sharples R. M., 1993, \name{\MNRAS}, \vol{263}, 560.

\ref Parry I. R., Sharples R. M., 1988, in \name{Fibre Optics in Astronomy,
Astronomical Soc. of the Pacific Conference Series Vol. 3},
ed. S.C. Barden, p. 93

\ref Peletier R. F., Valentijn E. A., Jameson R. F., 1990, \name{\AstrAst}, 233, 62

\ref Rocca-Volmerange, B., Guiderdoni, B., 1988, \name{\AstrAstSup}, 
\vol{75}, 93 (RVG)

\ref Rocca-Volmerange B., Guiderdoni B., 1990, \name{\MNRAS}, \vol{247}, 166

\ref Schneider D. P., Gunn J. E., Hoessel J. G., , 1983, \name{\ApJ}, \vol{268}, 476

\ref Songaila A., Cowie L. L., Hu E. M., Gardner J. P., 1994, \name{ApJ Suppl.}, \vol{94}, 461 

\ref Tonry J., Davis M., 1979, \name{\AstrJ}, \vol{84}, 1511

\ref Wynn C. G., Worswick S. P., 1988, \name{Observatory}, \vol{108}, 161

\ref Zucca E., Pozzetti L., Zamorani G., 1994, \name{\MNRAS}, \vol{269}, 953


\section*{APPENDIX: THE SPECTROSCOPIC SAMPLE}

This appendix presents details of our galaxy sample, listed
in Table A1.
The ID  numbers correspond to those in Table~4 of
Paper I. Note that, owing to an error, not all of Table~4 was
printed in Paper I; details of the omitted objects are available
on request from the authors. 

The spectral type is denoted by
a simple nomenclature where `E' and `A' respectively
refer to whether the
spectra have emission (predominantly [OII], $H\beta$ and [OIII])
and absorption features ($H+K$, $G$, $H\beta$--$H\theta$). Additionally
we use `ER' to refer to spectra with strong $H\alpha$ emission
where observable.
A selection of the survey spectra are shown in Figure~A1.

We give photometry both within 4-arcsec diameter apertures,
as in paper I, and also within 20 $h^{-1}$ kpc metric apertures.
For about 10\% of our objects, we were
unable to measure a big enough aperture at low redshift or small enough
at high redshift to reach 20 $h^{-1}$ kpc. For these objects we extrapolate
with the $r^{0.4}$ growth law -- typically the extrapolations are over
a small range of 20--50\% in diameter and thus deviations from the
$r^{0.4}$ law will be only at the $\ll0.1$ magnitude level. 

The last two columns give weights for incompleteness
as a function of $K$ and colour. The former was
obtained by the ratio of the raw number counts in the
sample to the mean count expected over 552 arcmin$^2$.
The latter was obtained by
dividing the $(R-K)$ -- $K$ colour--magnitude plane
into cells (using increments of 0.3 in $K$
and 0.5 in $R-K$) and comparing the populations of these
cells in our total and spectroscopic samples.
A colour-dependent weight was then deduced, normalised
to a mean of unity at each $K$. Where cell populations
were too low for meaningful statistics, a weight of
unity was assumed.

\vfill\eject
\strut
\vfill\eject

{

\onecolumn


\noindent
{\bf Table A1: The $K$-selected spectroscopic sample} 

\vfill

{\let\h=\hfil

\halign{\tabskip=1em
\h$#$& \h#\h & \h#\h & \h$#$\h & \h# & \h$#$\h & \h$#$\h & \h$#$\h & \h$#$\h & \h$#$\h & \h$#$\h \cr
\rm ID &\omit\h RA \quad (1950.0) \span\omit \quad DEC \h & z & Ty & K_{4''} & K_{20 h^{-1} \rm kpc} & R_{4''} & R_{20 h^{-1} \rm kpc} & w_K & w_{R-K} \cr
\noalign{\vglue 0.5truecm}
11 & 22 38 59.55 & +00 37 38.27 & 0.129 & A & 14.25 \pm 0.05 & 13.78 \pm 0.05 & 17.84 \pm 0.04 & 17.05 \pm 0.03 & 0.24 & 1.00 \cr 
18 & 22 38 41.69 & +00 31 49.68 & 0.384 & A & 16.43 \pm 0.10 & 16.44 \pm 0.16 & 20.39 \pm 0.05 & 20.08 \pm 0.04 & 3.93 & 0.81 \cr 
24 & 22 38 37.94 & +00 33 18.39 & 0.075 & E & 16.08 \pm 0.11 & 15.42 \pm 0.12 & 18.48 \pm 0.04 & 17.58 \pm 0.03 & 2.64 & 1.00 \cr 
31 & 22 38 43.84 & +00 33 55.83 & 0.666 & A & 15.90 \pm 0.09 & 15.75 \pm 0.12 & 20.20 \pm 0.05 & 20.13 \pm 0.04 & 2.09 & 1.00 \cr 
54 & 22 38 38.45 & +00 34 51.21 & 0.278 & EA & 16.96 \pm 0.17 & 16.65 \pm 0.21 & 20.68 \pm 0.06 & 20.15 \pm 0.05 & 3.17 & 1.33 \cr 
56 & 22 38 34.54 & +00 35 28.69 & 0.278 & EA & 16.74 \pm 0.15 & 16.35 \pm 0.19 & 19.87 \pm 0.05 & 19.48 \pm 0.04 & 2.69 & 0.98 \cr 
66 & 22 40 30.26 & +00 25 56.63 & 0.399 & EA & 16.92 \pm 0.18 & 16.65 \pm 0.22 & 20.79 \pm 0.05 & 20.52 \pm 0.04 & 3.17 & 1.29 \cr 
68 & 22 40 31.98 & +00 26 12.71 & 0.147 & EA & 16.25 \pm 0.10 & 15.97 \pm 0.18 & 19.00 \pm 0.04 & 17.92 \pm 0.03 & 2.64 & 1.00 \cr 
73 & 22 40 27.78 & +00 27 11.34 & 0.148 & A & 15.48 \pm 0.07 & 15.46 \pm 0.14 & 18.67 \pm 0.04 & 18.55 \pm 0.03 & 0.91 & 0.98 \cr 
77 & 22 40 02.69 & +00 21 34.23 & 0.160 & A & 16.31 \pm 0.12 & 15.88 \pm 0.16 & 19.70 \pm 0.04 & 19.18 \pm 0.03 & 3.93 & 1.21 \cr 
90 & 22 40 12.72 & +00 21 29.73 & 0.302 & A & 16.84 \pm 0.18 & 16.88 \pm 0.20 & 18.22 \pm 0.04 & 18.24 \pm 0.03 & 3.17 & 1.00 \cr 
91 & 22 40 09.02 & +00 21 37.83 & 0.191 & A & 15.83 \pm 0.09 & 15.54 \pm 0.14 & 19.00 \pm 0.04 & 18.34 \pm 0.03 & 2.09 & 1.00 \cr 
95 & 22 40 11.48 & +00 22 26.00 & 0.289 & A & 16.33 \pm 0.12 & 16.44 \pm 0.18 & 19.85 \pm 0.04 & 19.68 \pm 0.03 & 3.93 & 0.79 \cr 
96 & 22 40 13.93 & +00 22 32.01 & 0.225 & E & 16.04 \pm 0.10 & 15.98 \pm 0.15 & 21.81 \pm 0.10 & 21.63 \pm 0.09 & 2.64 & 1.00 \cr 
104 & 22 40 18.20 & +00 24 30.77 & 0.654 & A & 17.14 \pm 0.20 & 16.84 \pm 0.24 & 22.07 \pm 0.11 & 21.77 \pm 0.09 & 8.91 & 1.00 \cr 
106 & 22 40 18.74 & +00 24 58.85 & 0.801 & A & 16.22 \pm 0.10 & 16.16 \pm 0.10 & 21.70 \pm 0.09 & 21.63 \pm 0.06 & 2.64 & 1.00 \cr 
109 & 22 40 20.59 & +00 25 39.50 & 0.148 & EA & 15.42 \pm 0.06 & 14.96 \pm 0.08 & 19.38 \pm 0.04 & 18.76 \pm 0.03 & 0.91 & 0.86 \cr 
118 & 22 40 15.23 & +00 30 04.37 & 0.654 & A & 16.67 \pm 0.15 & 16.71 \pm 0.14 & 21.33 \pm 0.07 & 21.18 \pm 0.05 & 2.69 & 0.86 \cr 
120 & 22 40 19.71 & +00 30 14.37 & 0.210 & A & 15.69 \pm 0.08 & 15.51 \pm 0.12 & 19.00 \pm 0.05 & 18.42 \pm 0.04 & 0.83 & 0.98 \cr 
121 & 22 40 16.51 & +00 30 18.37 & 0.582 & A & 15.95 \pm 0.09 & 15.96 \pm 0.08 & 19.95 \pm 0.04 & 19.64 \pm 0.03 & 2.09 & 1.00 \cr 
132 & 22 39 58.28 & +00 24 46.24 & 0.409 & E & 16.82 \pm 0.15 & 16.68 \pm 0.19 & 20.04 \pm 0.05 & 19.88 \pm 0.04 & 3.17 & 0.98 \cr 
138 & 22 39 37.84 & +00 21 26.98 & 0.271 & EA & 16.11 \pm 0.10 & 15.91 \pm 0.13 & 20.12 \pm 0.05 & 19.53 \pm 0.04 & 2.64 & 1.00 \cr 
146 & 22 39 38.01 & +00 21 50.51 & 0.291 & EA & 16.75 \pm 0.15 & 16.58 \pm 0.20 & 19.93 \pm 0.05 & 19.34 \pm 0.03 & 2.69 & 0.98 \cr 
149 & 22 39 33.89 & +00 22 29.13 & 0.074 & ER & 15.03 \pm 0.06 & 14.47 \pm 0.11 & \hbox{No Data} & \hbox{No Data} & 0.63 & 1.00 \cr 
151 & 22 39 40.14 & +00 24 16.61 & 0.153 & EA & 15.56 \pm 0.07 & 15.55 \pm 0.14 & 18.49 \pm 0.04 & 17.82 \pm 0.03 & 0.83 & 1.02 \cr 
159 & 22 39 37.16 & +00 25 44.39 & 0.073 & A & 13.30 \pm 0.05 & 12.67 \pm 0.05 & 16.28 \pm 0.04 & 15.10 \pm 0.03 & 1.00 & 1.00 \cr 
164 & 22 39 38.59 & +00 28 33.25 & 0.408 & EA & 16.88 \pm 0.17 & 16.51 \pm 0.15 & 20.95 \pm 0.16 & 20.72 \pm 0.13 & 3.17 & 1.00 \cr 
169 & 22 39 32.42 & +00 31 27.18 & 0.150 & A & 15.73 \pm 0.10 & 15.31 \pm 0.10 & 18.94 \pm 0.04 & 18.30 \pm 0.03 & 0.83 & 0.98 \cr 
173 & 22 39 34.22 & +00 31 53.23 & 0.127 & A & 15.39 \pm 0.09 & 14.52 \pm 0.07 & 18.48 \pm 0.04 & 17.31 \pm 0.03 & 0.91 & 1.29 \cr 
181 & 22 39 31.48 & +00 25 00.98 & 0.332 & EA & 16.98 \pm 0.18 & 16.47 \pm 0.21 & 19.43 \pm 0.04 & 18.92 \pm 0.03 & 3.17 & 1.00 \cr 
190 & 22 39 05.99 & +00 22 58.60 & 0.387 & A & 15.93 \pm 0.09 & 15.95 \pm 0.12 & 19.82 \pm 0.04 & 19.67 \pm 0.03 & 2.09 & 1.00 \cr 
198 & 22 38 55.21 & +00 24 03.35 & 0.290 & A & 16.00 \pm 0.09 & 15.81 \pm 0.10 & 19.31 \pm 0.04 & 18.83 \pm 0.03 & 2.09 & 1.00 \cr 
199 & 22 38 56.90 & +00 24 14.64 & 0.388 & EA & 16.89 \pm 0.17 & 16.88 \pm 0.21 & 20.54 \pm 0.05 & 20.24 \pm 0.04 & 3.17 & 1.29 \cr 
212 & 22 39 00.35 & +00 29 45.16 & 0.179 & EA & 16.91 \pm 0.18 & 16.50 \pm 0.21 & 20.13 \pm 0.07 & 19.47 \pm 0.06 & 3.17 & 0.98 \cr 
220 & 22 38 47.57 & +00 26 21.46 & 0.777 & A & 16.78 \pm 0.15 & 16.67 \pm 0.15 & 21.50 \pm 0.08 & 21.37 \pm 0.07 & 3.17 & 0.86 \cr 
224 & 22 38 39.84 & +00 26 19.33 & 0.063 & A & 16.20 \pm 0.11 & 15.79 \pm 0.15 & 20.01 \pm 0.04 & 19.48 \pm 0.05 & 2.64 & 0.79 \cr 
226 & 22 38 39.41 & +00 26 32.98 & 0.358 & EA & 17.15 \pm 0.20 & 16.77 \pm 0.22 & 20.80 \pm 0.06 & 20.42 \pm 0.04 & 8.91 & 1.33 \cr 
229 & 22 38 39.37 & +00 27 32.27 & 0.503 & A & 16.62 \pm 0.13 & 16.39 \pm 0.15 & 20.33 \pm 0.05 & 19.76 \pm 0.05 & 2.69 & 0.81 \cr 
244 & 22 38 37.48 & +00 29 50.20 & 0.442 & A & 16.61 \pm 0.18 & 16.34 \pm 0.21 & 20.29 \pm 0.05 & 20.02 \pm 0.03 & 2.69 & 0.81 \cr 
250 & 22 38 53.87 & +00 20 58.55 & 0.658 & E & 16.52 \pm 0.17 & 16.41 \pm 0.22 & 22.22 \pm 0.12 & 22.15 \pm 0.09 & 2.69 & 1.00 \cr 
253 & 22 39 03.52 & +00 22 30.43 & 0.301 & A & 15.13 \pm 0.07 & 15.09 \pm 0.09 & 18.85 \pm 0.04 & 18.60 \pm 0.03 & 0.63 & 1.00 \cr 
259 & 22 38 50.30 & +00 29 11.81 & 0.276 & EA & 16.20 \pm 0.14 & 16.05 \pm 0.21 & 19.72 \pm 0.05 & 19.22 \pm 0.03 & 2.64 & 0.79 \cr 
265 & 00 52 10.66 & +00 15 35.34 & 0.207 & A & 14.73 \pm 0.05 & 14.22 \pm 0.06 & 18.03 \pm 0.04 & 17.45 \pm 0.03 & 1.26 & 1.00 \cr 
267 & 00 52 06.99 & +00 15 54.34 & 0.205 & A & 15.80 \pm 0.08 & 15.49 \pm 0.12 & 18.79 \pm 0.04 & 18.34 \pm 0.03 & 2.09 & 1.00 \cr 
281 & 00 52 29.84 & +00 16 48.97 & 0.210 & ER & 16.97 \pm 0.16 & 16.63 \pm 0.22 & 19.42 \pm 0.04 & 19.08 \pm 0.03 & 3.17 & 1.00 \cr 
285 & 00 52 34.76 & +00 20 19.43 & 0.067 & E & 15.56 \pm 0.08 & 15.01 \pm 0.13 & 18.14 \pm 0.04 & 17.39 \pm 0.03 & 0.83 & 1.02 \cr 
288 & 00 52 37.75 & +00 15 38.02 & 0.146 & ER & 15.57 \pm 0.07 & 14.97 \pm 0.10 & 18.60 \pm 0.04 & 17.62 \pm 0.03 & 0.83 & 0.98 \cr 
291 & 00 52 13.80 & +00 23 32.68 & 0.380 & E & 16.96 \pm 0.20 & 16.82 \pm 0.20 & 20.36 \pm 0.05 & 20.16 \pm 0.04 & 3.17 & 0.67 \cr 
293 & 00 52 15.23 & +00 23 49.49 & 0.377 & A & 15.52 \pm 0.08 & 15.28 \pm 0.07 & 18.76 \pm 0.04 & 18.50 \pm 0.03 & 0.83 & 0.98 \cr 
295 & 00 52 14.31 & +00 23 56.77 & 0.236 & A & 15.36 \pm 0.07 & 14.95 \pm 0.07 & 18.42 \pm 0.04 & 18.00 \pm 0.03 & 0.91 & 1.29 \cr 

}

\eject

\halign{\tabskip=1em
\h$#$& \h#\h & \h#\h & \h$#$\h & \h# & \h$#$\h & \h$#$\h & \h$#$\h & \h$#$\h & \h$#$\h & \h$#$\h \cr
\rm ID & \omit\h RA \quad (1950.0) \span\omit \quad DEC \h  & z & Ty & K_{4''} & K_{20 h^{-1} \rm kpc} & R_{4''} & R_{20 h^{-1} \rm kpc} & w_K & w_{R-K} \cr
\noalign{\vglue 0.5truecm}
297 & 00 52 04.75 & +00 20 52.75 & 0.067 & ER & 15.43 \pm 0.07 & 14.16 \pm 0.08 & 18.05 \pm 0.04 & 16.01 \pm 0.03 & 0.91 & 0.86 \cr 
300 & 00 52 09.76 & +00 21 30.25 & 0.087 & EA & 16.04 \pm 0.09 & 15.09 \pm 0.11 & 18.34 \pm 0.04 & 16.91 \pm 0.03 & 2.64 & 1.00 \cr 
302 & 00 52 08.35 & +00 21 48.30 & 0.086 & E & 14.18 \pm 0.05 & 13.48 \pm 0.06 & 16.95 \pm 0.04 & 16.02 \pm 0.03 & 0.37 & 1.00 \cr 
312 & 00 52 20.15 & +00 24 14.74 & 0.236 & A & 14.97 \pm 0.06 & 14.77 \pm 0.06 & 18.51 \pm 0.04 & 18.12 \pm 0.03 & 1.87 & 1.00 \cr 
313 & 00 52 17.60 & +00 24 20.68 & 0.235 & A & 16.25 \pm 0.10 & 16.07 \pm 0.18 & 19.63 \pm 0.05 & 19.30 \pm 0.04 & 2.64 & 1.21 \cr 
314 & 00 52 21.87 & +00 24 29.16 & 0.471 & A & 17.02 \pm 0.18 & 16.82 \pm 0.22 & 20.06 \pm 0.05 & 19.86 \pm 0.04 & 8.91 & 0.67 \cr 
316 & 00 52 25.49 & +00 20 11.44 & 0.676 & A & 16.80 \pm 0.13 & 16.71 \pm 0.18 & 21.50 \pm 0.09 & 21.42 \pm 0.07 & 3.17 & 0.86 \cr 
317 & 00 52 24.21 & +00 20 29.08 & 0.124 & A & 15.20 \pm 0.06 & 14.89 \pm 0.07 & 17.94 \pm 0.04 & 17.51 \pm 0.03 & 0.63 & 0.86 \cr 
319 & 00 52 25.52 & +00 20 39.10 & 0.546 & EA & 16.92 \pm 0.16 & 16.79 \pm 0.21 & 20.41 \pm 0.06 & 20.28 \pm 0.05 & 3.17 & 0.98 \cr 
320 & 00 52 24.81 & +00 20 45.78 & 0.124 & E & 16.63 \pm 0.12 & 16.09 \pm 0.20 & 18.79 \pm 0.04 & 17.67 \pm 0.03 & 2.69 & 1.00 \cr 
321 & 00 52 16.03 & +00 18 58.77 & 0.217 & EA & 16.50 \pm 0.13 & 16.48 \pm 0.21 & 20.14 \pm 0.05 & 19.98 \pm 0.04 & 3.93 & 0.81 \cr 
323 & 00 52 17.10 & +00 19 55.94 & 0.321 & E & 16.93 \pm 0.17 & 16.49 \pm 0.21 & 19.20 \pm 0.04 & 18.60 \pm 0.03 & 3.17 & 0.86 \cr 
329 & 00 52 27.57 & +00 21 23.25 & 0.153 & A & 15.75 \pm 0.10 & 15.53 \pm 0.18 & 19.33 \pm 0.04 & 18.81 \pm 0.03 & 0.83 & 1.00 \cr 
333 & 00 52 46.55 & +00 10 49.16 & 0.417 & A & 16.86 \pm 0.17 & 16.64 \pm 0.19 & 20.65 \pm 0.05 & 20.59 \pm 0.04 & 3.17 & 1.29 \cr 
334 & 00 52 47.79 & +00 10 52.72 & 0.192 & A & 15.83 \pm 0.08 & 15.62 \pm 0.12 & 19.80 \pm 0.04 & 18.48 \pm 0.03 & 2.09 & 1.00 \cr 
336 & 00 52 48.07 & +00 11 11.93 & 0.193 & EA & 16.57 \pm 0.13 & 16.48 \pm 0.18 & 19.99 \pm 0.04 & 19.63 \pm 0.03 & 2.69 & 1.19 \cr 
337 & 00 52 48.86 & +00 11 18.85 & 0.577 & A & 16.38 \pm 0.12 & 16.27 \pm 0.11 & 19.53 \pm 0.04 & 19.37 \pm 0.03 & 3.93 & 1.19 \cr 
340 & 00 52 48.71 & +00 11 51.46 & 0.636 & A & 15.59 \pm 0.07 & 15.47 \pm 0.07 & 19.86 \pm 0.04 & 19.71 \pm 0.03 & 0.83 & 1.00 \cr 
341 & 00 52 42.98 & +00 12 12.43 & 0.344 & EA & 16.66 \pm 0.14 & 16.43 \pm 0.14 & 19.82 \pm 0.04 & 19.57 \pm 0.03 & 2.69 & 0.98 \cr 
346 & 00 53 10.28 & +00 10 42.10 & 0.102 & A & 16.59 \pm 0.13 & 15.69 \pm 0.18 & 20.06 \pm 0.05 & 19.38 \pm 0.06 & 2.69 & 1.19 \cr 
349 & 00 53 11.74 & +00 11 38.97 & 0.154 & EA & 16.56 \pm 0.12 & 15.96 \pm 0.13 & 19.13 \pm 0.04 & 18.05 \pm 0.03 & 2.69 & 1.00 \cr 
354 & 00 53 12.64 & +00 05 48.91 & 0.153 & A & 14.29 \pm 0.05 & 13.66 \pm 0.05 & 17.80 \pm 0.04 & 17.00 \pm 0.03 & 0.24 & 1.00 \cr 
355 & 00 53 08.76 & +00 05 55.12 & 0.071 & E & 16.68 \pm 0.15 & 16.21 \pm 0.17 & 18.88 \pm 0.04 & 18.41 \pm 0.03 & 2.69 & 0.86 \cr 
356 & 00 53 09.54 & +00 06 05.00 & 0.209 & A & 15.08 \pm 0.06 & 14.84 \pm 0.08 & 18.27 \pm 0.04 & 17.87 \pm 0.03 & 0.63 & 1.00 \cr 
358 & 00 53 16.43 & +00 06 08.82 & 0.495 & A & 16.15 \pm 0.10 & 15.95 \pm 0.09 & 19.73 \pm 0.04 & 19.50 \pm 0.03 & 2.64 & 0.79 \cr 
360 & 00 52 48.99 & +00 09 26.06 & 0.673 & A & 17.24 \pm 0.21 & 17.20 \pm 0.27 & 22.00 \pm 0.11 & 21.96 \pm 0.08 & 8.91 & 1.00 \cr 
362 & 00 52 47.71 & +00 09 43.32 & 0.505 & EA & 17.10 \pm 0.18 & 17.11 \pm 0.20 & 20.18 \pm 0.04 & 20.09 \pm 0.03 & 8.91 & 0.67 \cr 
363 & 00 52 46.35 & +00 10 04.23 & 0.663 & A & 16.32 \pm 0.10 & 15.96 \pm 0.08 & 19.36 \pm 0.04 & 18.92 \pm 0.03 & 3.93 & 1.21 \cr 
364 & 00 52 47.10 & +00 10 19.78 & 0.431 & A & 16.58 \pm 0.12 & 16.46 \pm 0.13 & 20.14 \pm 0.04 & 19.92 \pm 0.03 & 2.69 & 0.81 \cr 
374 & 00 52 51.40 & +00 05 29.65 & 0.044 & EA & 16.92 \pm 0.18 & 15.69 \pm 0.20 & 19.04 \pm 0.04 & 17.53 \pm 0.03 & 3.17 & 0.86 \cr 
378 & 00 52 51.47 & +00 07 28.52 & 0.113 & EA & 17.10 \pm 0.20 & 16.21 \pm 0.19 & 19.70 \pm 0.04 & 18.81 \pm 0.03 & 8.91 & 1.00 \cr 
380 & 00 52 52.22 & +00 07 45.79 & 0.734 & A & 17.19 \pm 0.20 & 17.06 \pm 0.32 & 22.65 \pm 0.19 & 22.52 \pm 0.18 & 8.91 & 1.00 \cr 
385 & 00 52 59.56 & +00 09 44.62 & 0.045 & A & 14.35 \pm 0.05 & 12.70 \pm 0.05 & 16.82 \pm 0.04 & 15.14 \pm 0.03 & 0.24 & 1.00 \cr 
387 & 00 53 02.15 & +00 09 46.91 & 0.675 & A & 16.88 \pm 0.16 & 16.71 \pm 0.26 & 20.85 \pm 0.13 & 20.68 \pm 0.13 & 3.17 & 1.29 \cr 
392 & 00 53 07.44 & +00 07 15.48 & 0.153 & EA & 16.24 \pm 0.10 & 16.12 \pm 0.15 & 20.18 \pm 0.05 & 19.79 \pm 0.04 & 2.64 & 0.79 \cr 
400 & 00 53 11.97 & +00 13 25.19 & 0.192 & A & 17.21 \pm 0.19 & 16.81 \pm 0.26 & 20.46 \pm 0.06 & 20.06 \pm 0.06 & 8.91 & 0.67 \cr 
405 & 01 53 01.29 & +00 41 51.41 & 0.080 & ER & 15.73 \pm 0.09 & 15.04 \pm 0.09 & 18.50 \pm 0.04 & 17.19 \pm 0.03 & 0.83 & 1.02 \cr 
406 & 01 53 05.36 & +00 42 32.46 & 0.121 & A & 14.28 \pm 0.05 & 13.78 \pm 0.07 & 18.18 \pm 0.04 & 17.10 \pm 0.03 & 0.24 & 1.00 \cr 
411 & 01 53 28.63 & +00 46 32.36 & 0.085 & A & 14.43 \pm 0.06 & 13.86 \pm 0.07 & 17.39 \pm 0.04 & 16.73 \pm 0.03 & 0.24 & 1.00 \cr 
418 & 01 53 31.84 & +00 45 49.69 & 0.079 & A & 15.05 \pm 0.07 & 14.53 \pm 0.10 & 17.74 \pm 0.04 & 17.13 \pm 0.03 & 0.63 & 1.00 \cr 
422 & 01 53 21.37 & +00 46 21.86 & 0.113 & A & 15.66 \pm 0.07 & 15.48 \pm 0.10 & 19.15 \pm 0.04 & 18.94 \pm 0.04 & 0.83 & 0.98 \cr 
423 & 01 53 20.49 & +00 46 52.46 & 0.080 & A & 15.27 \pm 0.06 & 14.88 \pm 0.19 & 17.95 \pm 0.04 & 17.37 \pm 0.03 & 0.91 & 0.86 \cr 
425 & 01 53 24.68 & +00 44 39.93 & 0.080 & A & 14.68 \pm 0.05 & 13.76 \pm 0.07 & 17.51 \pm 0.04 & 16.44 \pm 0.03 & 1.26 & 1.00 \cr 
431 & 01 53 21.72 & +00 41 17.11 & 0.076 & ER & 15.67 \pm 0.09 & 15.11 \pm 0.21 & 18.13 \pm 0.04 & 16.76 \pm 0.03 & 0.83 & 1.00 \cr 
432 & 01 53 20.86 & +00 41 57.87 & 0.080 & ER & 15.42 \pm 0.08 & 14.67 \pm 0.09 & 18.04 \pm 0.04 & 17.19 \pm 0.03 & 0.91 & 0.86 \cr 
434 & 01 53 31.18 & +00 43 20.30 & 0.080 & A & 14.23 \pm 0.05 & 13.91 \pm 0.06 & 17.25 \pm 0.04 & 16.77 \pm 0.03 & 0.37 & 1.00 \cr 
454 & 01 53 13.88 & +00 43 58.12 & 0.376 & A & 16.82 \pm 0.13 & 16.66 \pm 0.21 & 20.28 \pm 0.05 & 20.17 \pm 0.04 & 3.17 & 0.98 \cr 
455 & 01 53 13.50 & +00 44 41.53 & 0.206 & A & 15.77 \pm 0.07 & 15.60 \pm 0.11 & 18.77 \pm 0.04 & 18.65 \pm 0.03 & 2.09 & 1.00 \cr 
457 & 01 53 14.85 & +00 45 01.44 & 0.551 & A & 17.10 \pm 0.17 & 16.93 \pm 0.27 & 20.83 \pm 0.06 & 20.66 \pm 0.04 & 8.91 & 1.33 \cr 
462 & 01 53 12.42 & +00 40 35.81 & 0.080 & A & 15.22 \pm 0.08 & 14.28 \pm 0.10 & 17.95 \pm 0.04 & 16.22 \pm 0.03 & 0.63 & 0.86 \cr 
469 & 01 52 48.85 & +00 39 51.71 & 0.554 & A & 15.57 \pm 0.09 & 15.49 \pm 0.10 & 18.50 \pm 0.04 & 18.39 \pm 0.03 & 0.83 & 1.02 \cr 
491 & 01 52 35.55 & +00 36 27.18 & 0.474 & A & 15.63 \pm 0.07 & 15.43 \pm 0.08 & 19.34 \pm 0.04 & 19.13 \pm 0.03 & 0.83 & 1.00 \cr 
502 & 01 52 45.28 & +00 22 33.32 & 0.403 & A & 16.15 \pm 0.10 & 15.90 \pm 0.11 & 19.93 \pm 0.04 & 19.61 \pm 0.03 & 2.64 & 0.79 \cr 
506 & 01 52 44.12 & +00 23 21.20 & 0.290 & A & 16.43 \pm 0.12 & 16.28 \pm 0.10 & 19.46 \pm 0.04 & 19.31 \pm 0.03 & 3.93 & 1.19 \cr 

}

\vfill\eject

\halign{\tabskip=1em
\h$#$& \h#\h & \h#\h & \h$#$\h & \h# & \h$#$\h & \h$#$\h & \h$#$\h & \h$#$\h & \h$#$\h & \h$#$\h \cr
\rm ID & \omit\h RA \quad (1950.0) \span\omit \quad DEC \h  & z & Ty & K_{4''} & K_{20 h^{-1} \rm kpc} & R_{4''} & R_{20 h^{-1} \rm kpc} & w_K & w_{R-K} \cr
\noalign{\vglue 0.5truecm}
510 & 01 52 52.08 & +00 29 46.18 & 0.088 & ER & 15.23 \pm 0.06 & 14.83 \pm 0.09 & 17.96 \pm 0.04 & 17.31 \pm 0.03 & 0.63 & 0.86 \cr 
511 & 01 52 49.82 & +00 29 47.50 & 0.471 & EA & 16.66 \pm 0.12 & 16.56 \pm 0.20 & 20.48 \pm 0.05 & 20.17 \pm 0.03 & 2.69 & 1.29 \cr 
517 & 01 52 45.72 & +00 27 46.09 & 0.131 & ER & 14.85 \pm 0.05 & 14.40 \pm 0.07 & 18.22 \pm 0.04 & 17.51 \pm 0.03 & 1.87 & 1.00 \cr 
520 & 01 52 51.65 & +00 26 07.60 & 0.252 & EA & 16.63 \pm 0.12 & 16.23 \pm 0.17 & 20.03 \pm 0.05 & 19.58 \pm 0.04 & 2.69 & 1.19 \cr 
521 & 01 52 50.02 & +00 26 30.46 & 0.140 & EA & 16.68 \pm 0.12 & 16.24 \pm 0.21 & 18.97 \pm 0.04 & 18.16 \pm 0.03 & 2.69 & 0.86 \cr 
525 & 01 52 55.08 & +00 25 07.90 & 0.339 & EA & 17.00 \pm 0.18 & 16.77 \pm 0.18 & 20.19 \pm 0.05 & 19.97 \pm 0.04 & 3.17 & 0.67 \cr 
540 & 01 52 27.74 & +00 24 32.90 & 0.190 & A & 15.63 \pm 0.07 & 15.27 \pm 0.08 & 18.58 \pm 0.04 & 17.90 \pm 0.03 & 0.83 & 1.02 \cr 
558 & 01 52 41.29 & +00 28 27.72 & 0.119 & A & 15.43 \pm 0.07 & 14.25 \pm 0.07 & 17.80 \pm 0.04 & 16.48 \pm 0.03 & 0.91 & 1.00 \cr 
561 & 01 52 39.81 & +00 29 05.68 & 0.113 & A & 16.27 \pm 0.10 & 15.43 \pm 0.14 & 18.78 \pm 0.04 & 18.07 \pm 0.03 & 3.93 & 1.00 \cr 
563 & 01 52 36.76 & +00 25 14.35 & 0.080 & ER & 14.40 \pm 0.05 & 13.73 \pm 0.06 & 18.09 \pm 0.04 & 16.73 \pm 0.03 & 0.24 & 1.00 \cr 
564 & 01 52 37.05 & +00 25 30.23 & 0.339 & A & 16.92 \pm 0.16 & 16.44 \pm 0.21 & 19.70 \pm 0.04 & 19.17 \pm 0.03 & 3.17 & 1.00 \cr 
568 & 01 52 24.55 & +00 31 17.28 & 0.156 & A & 14.38 \pm 0.05 & 13.99 \pm 0.07 & 18.04 \pm 0.04 & 17.56 \pm 0.03 & 0.24 & 1.00 \cr 
571 & 01 52 38.20 & +00 39 55.06 & 0.169 & A & 15.20 \pm 0.08 & 14.78 \pm 0.11 & 18.50 \pm 0.04 & 18.20 \pm 0.03 & 0.63 & 1.29 \cr 
574 & 01 52 46.01 & +00 32 37.12 & 0.286 & A & 15.16 \pm 0.06 & 14.72 \pm 0.07 & 18.78 \pm 0.04 & 18.23 \pm 0.03 & 0.63 & 0.86 \cr 
576 & 01 52 44.69 & +00 31 30.01 & 0.287 & A & 15.42 \pm 0.08 & 15.34 \pm 0.12 & 19.32 \pm 0.04 & 18.48 \pm 0.03 & 0.91 & 0.86 \cr 
1445 & 13 42 05.07 & +00 09 32.74 & 0.255 & A & 17.19 \pm 0.27 & 16.89 \pm 0.35 & 20.30 \pm 0.05 & 20.00 \pm 0.04 & 8.91 & 0.67 \cr 
1446 & 13 42 08.32 & +00 09 56.45 & 0.370 & A & 16.45 \pm 0.15 & 16.24 \pm 0.14 & 20.52 \pm 0.05 & 20.31 \pm 0.04 & 3.93 & 1.00 \cr 
1450 & 13 42 05.43 & +00 10 26.78 & 0.430 & E & 16.68 \pm 0.18 & 16.50 \pm 0.30 & 20.54 \pm 0.05 & 20.36 \pm 0.04 & 2.69 & 1.29 \cr 
1459 & 13 42 03.58 & +00 05 40.13 & 0.408 & EA & 16.50 \pm 0.14 & 16.31 \pm 0.12 & 20.15 \pm 0.05 & 19.96 \pm 0.04 & 2.69 & 0.81 \cr 
1550 & 13 41 54.03 & +00 05 02.38 & 0.088 & A & 14.19 \pm 0.05 & 13.64 \pm 0.06 & 17.01 \pm 0.03 & 16.18 \pm 0.03 & 0.37 & 1.00 \cr 

}

}

}  

\vfill\eject

\section*{FIGURE CAPTIONS}

\noindent
{\bf Figure 1}\quad
(a) The $R-K$ vs $K$ colour-magnitude plane for our survey.  The open
circles show all the data, with spectroscopic sample members being
indicated by solid points.  Data from both March and October fields are
included, although a correction to the former is made so that the
magnitudes refer to 4-arcsec diameter apertures (see section 3).
Brightwards of the spectroscopic selection at $K=17.25$, the sampling
of colour at fixed $K$ is very close to uniform.
\strut\hfill\break
(b) The same as panel (a), but now the points repesent all galaxies,
the open circles show the 16 unidentified spectra which are brighter
than the appropriate optical completeness limits defined for the
spectroscopic runs, and the solid circles show the 8 identified
spectrscopic galaxies which lie {\em fainter}
 than the completeness limits. These are almost all within 0.5 mag. of
the relevant optical limit; our assumption is that these are merely
slightly degraded versions of our successful spectra, and that their
omission does not bias the results.

\strut\hfill\break
\noindent
{\bf Figure 2}\quad
Redshift against $K$ magnitude for the spectroscopic sample.
Note that the sample probes to substantial redshifts $z_{\rm max}=0.8$,
and that the high-redshift bins contain a number of relatively
bright galaxies, as bright at $K\simeq 15.5$ at $z=0.6$.

\strut\hfill\break
\noindent
{\bf Figure 3}\quad
The difference between our published 4 arcsecond $K$ magnitudes
and the new direct determinations of the magnitudes in
a 20 $h^{-1}$ Mpc aperture. The error bars are those
for the larger aperture measurement. The solid line shows the behaviour
expected for the adopted $r^{0.4}$ growth curve. This simple
a priori model is an excellent fit to the data, and
clearly introduces systematic errors no larger than about 0.1 mag.

\strut\hfill\break \noindent
{\bf Figure 4}\quad The $R-K$ vs $z$
colour-redshift distribution. The different lines show the loci of the
old Burst model (providing a red envelope for ellipticals) and Hubble
Sc and Im types from Rocca-Volmerange \& Guiderdoni (1988). Also shown
for comparison (dashed) is the old Burst model from Bruzual \& Charlot (1993).

\strut\hfill\break
\noindent
{\bf Figure 5}\quad
The $K$-band number counts, normalized to the usual
Euclidean slope. The smooth analytic fit described in the
text is statistically consistent with all the measurements,
showing that the counts are well determined.
All data have been corrected to
4-arcsec diameter aperture magnitudes, as described in the text.

\strut\hfill\break
\noindent
{\bf Figure 6}\quad
The `completeness' of our spectroscopic sample, expressed
as a ratio of the observed number of galaxies in a bin
to the number predicted by our count fitting formula
for 552 arcmin$^2$.
We assume that the redshift distribution at given $K$
is unbiased in our sample, and that the sampling fraction
shown here may be used to correct our sample to be
representative of one complete to $K=17.25$.
Note that the apparently unphysical values exceeding
unity are reasonable: some bright galaxies
were inadvertently selected as astrometric reference
`stars', so these bins are biased high.

\strut\hfill\break
\noindent
{\bf Figure 7}\quad
The histograms of redshift for our data. The raw numbers
are shown in (a) and (b) gives the result after weighting galaxies
to allow  for $K$-dependent sampling.

\strut\hfill\break
\noindent
{\bf Figure 8}\quad
The K-corrections for the Bruzual \& Charlot models. The
different lines show the behaviour for a delta-function
burst of star formation at different ages, from 1 to 10
Gyr. We shall use the 5 Gyr model as the default K-correction.
The satisfyingly near-universal predicted spectral
shape in the near-infrared is well evident in this plot.
Also shown (dashed) is the UV-hot elliptical model of
Rocca-Volmerange \& Guiderdoni (1988).

\strut\hfill\break
\noindent
{\bf Figure 9}\quad The redshift-magnitude data of figure 2,
translated to the redshift-absolute magnitude plane.  There is a
smooth increase of the maximum luminosity sampled with redshift.
However, in order to determine whether this corresponds to an
increasing characteristic luminosity, a full luminosity function
analysis is required to take account of the sampling volumes as a
function of redshift.
The different symbols correspond to three equal classes of
restframe colour: filled circles denote E/Sa; open circles Sb;
crosses Sc/Im. Note the fainter characteristic luminosity
of the last class.

\strut\hfill\break
\noindent
{\bf Figure 10}\quad
The luminosity function results, expressed in two ways:
(a) the binned luminosity function; (b) the cumulative
luminosity density, assuming $M_K(\odot)=3.4$. Both
these methods make the point that there is little
evidence for evolution in the luminosity function out to
$z=0.4$. At $0.4<z<0.8$, however, the characteristic
luminosity is 0.5 -- 1 mag. brighter, with some suggestion
that the characteristic density has declined.

\strut\hfill\break
\noindent
{\bf Figure 11}\quad
The $K$-band number counts, normalized to the usual
Euclidean slope. The plotting symbols have the same 
meanings as in Figure~5.
The various solid lines show different
models based on our low-redshift luminosity function
results, all assuming $\Omega=1$. Panel (a) shows that 
no evolution fits the data rather well, in contrast to
our earlier predictions based on a brighter $L^*$ and
lower $\phi^*$. Panel (b) shows that
pure luminosity evolution exceeds the faint counts.
Luminosity evolution with declining normalization at high redshift
fits better, but the predicted counts are too low in the faintest bins. 
Including a (non-evolving) dwarf component to the local LF as in
Driver et al. (1994) provides a good fit.

\bigskip
\noindent
{\bf Figure A1}\quad
Plots of a random selection of spectra from the survey.
The positions of standard spectral features are indicated for
the adopted redshift.

\strut\hfill\break

\vfill\eject

\bye   

\input psfig

\def\dofig#1{ \eject\psfig{file=fig#1.ps,width=\hsize,rheight=0pt,silent=} 
              \vfill
              \line{\hfill\bf Fig.~#1\hfill}
}

\dofig{1}
\dofig{2}
\dofig{3}
\dofig{4}
\dofig{5}
\dofig{6}
\dofig{7}
\dofig{8}
\dofig{9}
\dofig{10}
\dofig{11}
\eject
\hglue -23pt \psfig{file=figA1.ps,width=522pt,height=657pt,rwidth=0pt,silent=} 
\vfill
\line{\hfill\bf Fig.~A1\hfill}
\bye